\documentclass[aps,preprint,floatfix,nofootinbib,showpacs]{revtex4-1}
\pdfoutput=1
\usepackage{dcolumn}
\usepackage{bm}
\usepackage{graphicx}
\usepackage{amssymb,amsmath}
\usepackage{multirow}
\usepackage{units,changes}
\usepackage{color,url}
\usepackage{tabu}
\usepackage{array}
\usepackage[colorlinks=true,urlcolor=blue,anchorcolor=blue
,citecolor=blue,filecolor=blue,linkcolor=blue,menucolor=blue
,linktocpage=true,pdfproducer=medialab,pdfa=true]{hyperref}
\usepackage{colordvi}
\usepackage{subcaption}
\def\beqa{\begin{eqnarray}}
\def\eeqa{\end{eqnarray}}

\begin{document}

\title{ Crossing two-component dark matter models and implications for 511 keV 
$\gamma$-ray and XENON1T excesses}
\def\slash#1{#1\!\!\!/}

\author{P. Ko$^1$, Chih-Ting Lu$^1$, Ui Min$^2$}
\affiliation{
 $^1$ School of Physics, KIAS, Seoul 130-722, Republic of Korea \\ 
 $^2$ Department of Physics, Korea Advanced Institute of Science and Technology,291 Daehak-ro, Yuseong-gu, Daejeon 34141, Republic of Korea \\
}
\date{\today}

\begin{abstract}

We study scalar and fermionic crossing two-component dark matter (C2CDM) models with 
$ U(1)_X $ dark gauge symmetry. This $ U(1)_X $ gauge symmetry is broken via the dark Higgs mechanism of a dark Higgs field $\Phi$ and the dark photon becomes massive. On the other hand, the same dark Higgs field $\Phi$ also serves as a bridge between each component of two DM sectors such that the dark flavor-changing neutral current (DFCNC) interaction between them is induced. Moreover, the mass splitting between these two DM sectors is generated after $ U(1)_X $ gauge symmetry breaking by nonzero $\langle \Phi \rangle$. 
We discuss the stability of DM candidates and allowed parameter space from the relic 
density and other constraints in C2CDM models. Some novel signatures with displaced 
vertices at Belle II are studied.
Finally, the possible explanations of 511 keV $\gamma$-ray line and XENON1T excess 
in C2CDM models are discussed. 
 
\end{abstract}

\maketitle

\section{Introduction}

The particle nature of dark matter (DM) and its interactions with standard model (SM) 
particles and with itself are still puzzles. 
Until now there are only some gravitational evidences for  DM~\cite{Frenk:2012ph}. 
Therefore, exploring the properties of DM is an important mission. 
Among various DM models, the weakly interacting massive particles (WIMPs) are popular 
DM candidates.  And there are many ways to search for WIMPs including DM direct 
detection, DM indirect detection and collider experiments as shown in Ref.~\cite{Bauer:2017qwy,Arcadi:2017kky} and references therein.   
However, there are no concrete evidences for DM from these searches.  
Especially, the DM direct 
detection experiments provide the most stringent constraints. Taking $ M_{DM} = 100 $ GeV, the spin independent (spin dependent) DM and nuclear scattering cross section is highly 
restricted : 
\begin{center}
 $\sigma_{\text{SI}} \lesssim 10^{-46} \text{cm}^2$ ~\cite{XENON:2018voc,PandaX-4T:2021bab} \quad and \quad  
 $\sigma_{\text{SD}} \lesssim 2\times 10^{-41} \text{cm}^2$ ~\cite{XENON:2019rxp}.
\end{center}
Therefore, recent interests for DM tend to move to much heavier or lighter DM candidates
~\cite{Baek:2013dwa,Kim:2019udq,Baker:2019ndr,Kramer:2020sbb,Knapen:2017xzo,Lin:2019uvt}  
\footnote{Another possible WIMP scenario that is still viable 
is the so-called secluded case~\cite{Pospelov:2007mp,Pospelov:2008jd,Pospelov:2008zw}, where DM pair annihilations occur mainly into a pair of 
mediators lighter than DM, and then those mediators decay into the SM particles.  
The galactic center (GC) $\gamma$-ray excess \cite{Slatyer:2021qgc,Cholis:2021rpp} can be easily accommodated in this scenario \cite{Ko:2014gha,Baek:2014kna,Ko:2014loa,Ko:2015ioa}.}.

The DM candidates can be either stable or long-lived. If DM candidate is long-lived, its lifetime is forced to be much longer than the age of universe~\cite{Slatyer:2017sev}. On the other hand, if the DM candidates are stable, we may need extra exact symmetries to stabilize them. The discrete symmetries, like R-parity in MSSM~\cite{Jungman:1995df}, are popular choices to stabilize the DM candidates. However, once we take into account the quantum gravity effect, directly applying discrete or global continuous internal symmetries for the stability of DMs is problematic \cite{Baek:2013qwa}. It may induce the dimension five effective operators which will make DM candidates decay too fast unless it is very light  \cite{Baek:2013qwa}.  Therefore, imposing the local gauge symmetries for DM models 
is a better choice for the DM stability issue for internal discrete  symmetry~\cite{Krauss:1988zc}. For the case of continuous dark gauge symmetry 
stabilizing DM, one can find out some models and related discussions in Refs. 
\cite{Baek:2013qwa,Baek:2013dwa,Ko:2016fcd}. 

The DM models with $ U(1)_X $ gauge symmetry are attractive for phenomenological studies~\cite{Holdom:1985ag,Dienes:1996zr,Pospelov:2007mp,Chun:2010ve,Bauer:2018egk}. However, for the fermionic DM, its mass cannot be less than about $\sim 10 $ GeV from 
the CMB constraint if its dominant annihilation processes are in the $s$-wave~
\cite{Hutsi:2011vx,Ade:2015xua}. If we consider only the dark photon mediator, 
$A'$, in the relic density calculations, the fermionic DM pair annihilate 
through virtual $A'$ exchange in the $s$-channel will be in the $s$-wave, thereby 
stringently constrained by the CMB data. Unlike the fermionic DM, the dominant annihilation processes for the scalar DM with dark photon mediator is in the $p$-wave such that 
the sub-GeV scalar DM is still allowed and it only suffers from BBN constraint~\cite{Depta:2019lbe,Krnjaic:2019dzc} for its mass larger than about $ 10 $ MeV. 

However, the story can change completely if one includes dark Higgs boson, $\phi$, 
which should be generically present when dark photon gets massive by dark Higgs mechanism. In this case, new channels involving light dark Higgs boson can appear 
in the DM pair annihilations \cite{Baek:2020owl}: 
$X X^\ast \rightarrow A' \phi$ (scalar DM) or $\psi \bar\psi \rightarrow \phi\phi$ 
(fermionic DM), which are all in the $p$-wave.
Other dangerous $s$-wave contributions can be kinematically forbidden for a suitable 
choice of mass spectra in the dark sector \footnote{ 
There are other various DM physics topics where dark Higgs boson plays important roles, 
including restoration of unitarity at high energy scale: GC $\gamma$-ray excess 
\cite{Ko:2014gha,Ko:2014loa,Baek:2014kna,Ko:2015ioa},  DM bound state formations affecting relic density calculation \cite{Ko:2019wxq}, DM searches at high energy 
colliders \cite{Ko:2016xwd,Kamon:2017yfx,Baek:2015lna,Ko:2016ybp,Dutta:2017sod,Ko:2018mew}, and Higgs-portal assisted Higgs inflation \cite{Kim:2014kok}. }. 
Dark Higgs boson plays a crucial role 
for the light DM scenario to be thermal WIMP DM in a way consistent with 
CMB and BBN bounds. In short, DM phenomenology with massive dark photon 
can not be complete without including dark Higgs boson.

There are also many other proposals to alleviate the CMB constraint for the mass of 
fermionic DM to sub-GeV. For example, the asymmetric DM~\cite{Baldes:2017gzu}, 
inelastic DM~\cite{Izaguirre:2015zva}, and freeze-in mechanism~\cite{Bernal:2015ova} scenarios, etc. 
On the other hand, the dark sector may include plenty of new particle species and 
parts of them are stable or long-lived as our visible world~
\cite{Hur:2007ur,Ko:2010at,Baek:2013dwa,Petraki:2014uza,Ko:2014bka,Ko:2016fcd,Aoki:2016glu,Yaguna:2019cvp}. Moreover, this kind of multi-component DM models are interesting to generate some novel phenomena which are distinguishable from single-component DM models~\cite{Zurek:2008qg,Profumo:2009tb,Aoki:2012ub,Aoki:2018gjf}.

In this work, we study a type of two-component DM models with $ U(1)_X $ dark gauge symmetry which includes the crossing effect between each component. We call it as the crossing two-component dark matter (C2CDM) and consider both scalar and fermionic DM scenarios. 
We emphasize the function of new complex scalar dark Higgs field $\Phi$ in our C2CDM 
models is threefold. First, the $ U(1)_X $ dark gauge symmetry is broken via a dark Higgs mechanism by the nonzero VEV of $\Phi$, and the dark photon becomes massive. 
Second, the same dark Higgs field $\Phi$ also serves as a bridge between each 
component of two DM sectors such that dark flavor-changing neutral current (DFCNC) 
interaction between them is induced. Third, the mass splitting between these two DM particles is generated after $ U(1)_X $ gauge symmetry is broken. 
As we know, the FCNC interactions in visible sectors are already suffered from 
severe constraints and make them become rare processes~\cite{Glashow:1970gm}. 
It would be interesting to explore the situation for the DFCNC interactions~\cite{Ibe:2018juk,Ibe:2018tex,Choi:2020ysq,Herms:2021fql}. 
We found these DFCNC interactions can generate novel signatures at collider and fixed 
target experiments which cannot be produced from single-component DM or inelastic DM models with $ U(1)_X $ dark gauge symmetry. 
We take the Belle II experiment as an example to display the search strategy for these 
novel signatures.

On the other hand, there are still some  hints for DM which are still controversial however. For the DM indirect detections, the 3.5 keV line~\cite{Silich:2021sra,Slatyer:2021qgc}, 511 keV line~\cite{Kierans:2019aqz,Ema:2020fit,Keith:2021guq}, antiproton excess~\cite{Cuoco:2019kuu,Cholis:2019ejx} and galactic center GeV excess~\cite{Slatyer:2021qgc,Cholis:2021rpp}  still need to be further confirmed if these observations really come from the DM annihilations/decays or other astrophysics objects.  
For the DM direct detection, the recent XENON1T excess~\cite{XENON:2020rca} also 
catch our attention and it's still a puzzle if this excess comes from DM or not. 
In principle, all of the above puzzles cannot be interpreted in a single DM model. 
Therefore, we only discuss some possible explanations for 511 keV $\gamma$-ray line 
and XENON1T excess in C2CDM models in this work, mostly focusing on the 
light WIMP scenarios with $M_{\rm DM} \lesssim {\cal O}(1)$ GeV. 

The organization of this paper is as follows. We first show scalar and fermionic C2CDM models with $U(1)_X$ gauge symmetry in Sec.~\ref{Sec:Model}. The stability of DM candidates, relic density and DM direct detection are discussed in Sec.~\ref{Sec:stability}. 
We then study all other relevant constraints in C2CDM models in Sec.~\ref{Sec:constraints}. The novel dilepton displaced vertex signatures at Belle II for C2CDM models are studied in 
Sec.~\ref{Sec:BelleII}. The possible explanations of 511 keV galactic line and XENON1T excess in C2CDM models are discussed in Sec.~\ref{Sec:511X1T}. Finally, we conclude 
our studies in Sec.~\ref{Sec:Conclusion}. 
Some supplemental formulae for Sec.~\ref{Sec:Model} and~\ref{Sec:stability} are 
displayed in the Appendices~\ref{Sec:App1} and~\ref{Sec:App2}.

\section{The models}\label{Sec:Model} 

We display both scalar and fermionic C2CDM models with $ U(1)_X $ dark gauge 
symmetry in this section. 
We especially show the role of complex scalar dark Higgs  field $\Phi$ which provides 
the mass to dark photon after the spontaneous symmetry breaking (SSB) with nonzero VEV $\langle \Phi \rangle$, and triggers the mixing and the mass splitting between two-component DM sectors.  

\subsection{Scalar C2CDM models}\label{Sec:SC2CDM}

\begin{table}[t!]
  \begin{tabular}{l @{\extracolsep{0.2in}} r r r}
   \hline
Fields            & $ X_1 $ & $ X_2 $ & $ \Phi $ \\ \hline
$ U(1)_X $ charge & $ 1 $   & $ q_X $ & $ 1-q_X $ \\ \hline 
    \end{tabular}
    \caption{\small
    The associated $ U(1)_X $ charges for fields in scalar C2CDM models.}
\label{Tab:U1scoup}
\end{table}

We define three SM singlet complex scalar fields, $ X_1 $, $ X_2 $ and $ \Phi $ with $ U(1)_X $ charges in Table.~\ref{Tab:U1scoup}. We assume $ q_X $ is a real number and $ q_X\neq 0, 1 $ such that all $ X_1 $, $ X_2 $ and $ \Phi $ are charged under $ U(1)_X $. On the other hand, in order to simplify our discussion, we don't consider $ q_X = -1, -2, -3 $ which will induce $ \Phi^{\ast}X^2_1 $, $ \Phi X^2_2 $, $\Phi^{\ast}X^3_1$, $X^2_1 X_2$ and $X^3_1 X_2$ operators as in~\cite{Baek:2014kna,  Baek:2020owl,  Kang:2021oes,  Choi:2021yps}. 
These operators need special treatments or care.  
Notice all SM fields don't carry the $ U(1)_X $ charge in scalar C2CDM models.

The scalar part of the renormalizable and gauge invariant Lagrangian density is 
\begin{equation}
{\cal L}_{scalar} = |D_{\mu}H|^2 + |D_{\mu}\Phi|^2 + |D_{\mu}X_{1}|^2 + |D_{\mu}X_{2}|^2 -V(H,\Phi,X_1,X_2) 
\label{eq:Ls}
\end{equation} 
with 
\begin{align} 
& D_{\mu}H = (\partial_{\mu}+i\frac{g}{2}\sigma_{a}W^{a}_{\mu}+i\frac{g^{\prime}}{2}B_{\mu})H, 
\nonumber  \\ & 
D_{\mu}\Phi = (\partial_{\mu}+ig_X (1-q_X) C_{\mu})\Phi,
\nonumber  \\ & 
D_{\mu}X_{1} = (\partial_{\mu}+ig_X C_{\mu})X_{1},
\nonumber  \\ & 
D_{\mu}X_{2} = (\partial_{\mu}+ig_X q_X C_{\mu})X_{2},
\label{eq:covariant1}
\end{align}
where $ H $ is the SM scalar doublet and $ W^i_{\mu} $, $ B_{\mu} $, and $ C_{\mu} $ are gauge potentials of $SU(2)_L$, $U(1)_Y$ and $U(1)_X$ with gauge couplings $g$, $g'$ and $g_X$, respectively. The scalar potential in Eq.(\ref{eq:Ls}) is 
\begin{align} 
V(H,\Phi,X_1,X_2) = & 
-\mu^2_H H^{\dagger}H +\lambda_H (H^{\dagger}H)^2 -\mu^2_{\Phi}\Phi^{\ast}\Phi +\lambda_{\Phi}(\Phi^{\ast}\Phi)^2 
\nonumber  \\ & 
+\mu^2_{X_1} X^{\ast}_1 X_1 +\mu^2_{X_2} X^{\ast}_2 X_2 +\lambda_{X_1}(X^{\ast}_1 X_1)^2 +\lambda_{X_2}(X^{\ast}_2 X_2)^2 
\nonumber  \\ & 
+\lambda_{H\Phi}(H^{\dagger}H)(\Phi^{\ast}\Phi) +\lambda_{X_1 X_2}(X^{\ast}_1 X_1)(X^{\ast}_2 X_2) 
\nonumber  \\ & 
+\left[ \lambda_{HX_1}(H^{\dagger}H) +\lambda_{\Phi X_1}(\Phi^{\ast}\Phi)\right] X^{\ast}_1 X_1 
\nonumber  \\ & 
+\left[ \lambda_{HX_2}(H^{\dagger}H) +\lambda_{\Phi X_2}(\Phi^{\ast}\Phi)\right] X^{\ast}_2 X_2 
\nonumber  \\ &
+(\mu_{X_1 X_2 \Phi}X_1 X^{\ast}_2 \Phi^{\ast} +H.c.). 
\label{eq:potentail1}
\end{align}  
Note the last term in Eq.(\ref{eq:potentail1}) can trigger the mixing between $ X_1 $ and $ X_2 $ after the SSB of $ \Phi $, which is a  feature of this model with crossing two components. 
If we changed the $ U(1)_X $ charge assignment in Table~\ref{Tab:U1scoup} such that the 
last term in Eq.(\ref{eq:potentail1}) were not allowed, this model would turn to the simple 
$ U(1)_1\times U(1)_2 $ two-component scalar DM model. 
That is the reason why we call it as the "crossing" two-component DM model.

We then expand $H$ and $\Phi$ fields around the vacuum with the unitary gauge,
\begin{equation}
H =  
\left(
\begin{tabular}{c}
$0$
\\
$\frac{1}{\sqrt{2}}(v + h)$
\end{tabular}
\right)
\;\;\; , \;\;\; 
\Phi = \frac{1}{\sqrt 2}\left( v_X + h_X \right).
\label{eq:expand}
\end{equation}
The mass eigenstates $h_1$ and $h_2$ can be defined from the interaction eigenstates $h$ and $h_X$ as
\begin{equation}
h_1 = h \cos \theta_h + h_X \sin \theta_h, \ h_2=  -h \sin \theta_h + h_X \cos \theta_h ,
\end{equation}
and their mass eigenvalues are
\begin{equation}
m^2_{h_1,h_2} = \lambda v^2 + \lambda_\Phi v^2_X \pm \sqrt{\left( \lambda v^2 - \lambda_\Phi v^2_X \right)^2 +\lambda^2_{H\Phi} v^2 v^2_X}, 
\end{equation} 
where the mixing angle $\theta_h$ satisfies 
\[
\tan 2\theta_h = \frac{\lambda_{H\Phi} v v_X}{\lambda v^2 - \lambda_\Phi v^2_X} .
\]
We assign $h_1$ as the SM-like Higgs boson with mass of $ 125 $ GeV.

The mass term of DM sectors can be written as 
\begin{equation}
{\cal L}_{DM,mass} =  - 
\left( X^{\ast}_1 X^{\ast}_2 \right)   
\left(
\begin{tabular}{cc}
$ \mu^2_{X_1} + (\lambda_{HX_1}v^2 +\lambda_{\Phi X_1}v^2_X)/2$ & $\mu_{X_1 X_2 \Phi} v_X /\sqrt{2}$ 
\\
$\mu_{X_1 X_2 \Phi} v_X /\sqrt{2}$ & $ \mu^2_{X_2} + (\lambda_{HX_2}v^2 +\lambda_{\Phi X_2}v^2_X)/2$ 
\end{tabular}
\right) 
\left(
\begin{tabular}{c}
$X_1$
\\
$X_2$
\end{tabular}
\right)
\label{eq:mass1}
\end{equation}
and the mass eigenstates $X'_1$ and $X'_2$ can be rotated from the interaction eigenstates $X_1$ and $X_2$ as
\begin{equation}
X'_1 = X_1 \cos \theta_X - X_2 \sin \theta_X, X'_2 = X_1 \sin \theta_X + X_2 \cos \theta_X ~,
\end{equation}
where the mixing angle $\theta_X$ satisfies 
\begin{equation}
\tan 2\theta_X = \frac{\sqrt{2} \mu_{X_1 X_2 \Phi} v_X }{ \mu^2_{X_2} + \frac{1}{2} \lambda_{HX_2} v^2 + \frac{1}{2} \lambda_{\Phi X_2} v^2_X -\mu^2_{X_1} - \frac{1}{2} \lambda_{HX_1} v^2 - \frac{1}{2} \lambda_{\Phi X_1} v^2_X }~.  
\end{equation}
Note that the mass eigenstates $X'_1$ and $X'_2$ are linear combinations of 
two states $X_1$ and $X_2$ with different $U(1)_X$ charges after spontaneously breaking 
of $U(1)_X$ gauge symmetry. This could lead to interesting phenomena in the DM 
self-scattering and a possibility of bound state formations, especially when one considers 
the contributions of dark Higgs boson as well \cite{Ko:2019wxq}.
Their mass eigenvalues can be solved as
\begin{equation}
\begin{split}
& M^2_{X'_1, X'_2} = \frac{1}{2} \left(  \mu^2_{X_1} + \frac{1}{2} \lambda_{HX_1} v^2 + \frac{1}{2} \lambda_{\Phi X_1} v^2_X +\mu^2_{X_2} + \frac{1}{2} \lambda_{HX_2} v^2 + \frac{1}{2} \lambda_{\Phi X_2} v^2_X  \right) 
\\ & \mp \frac{1}{2} \sqrt{ \left(  \mu^2_{X_2} + \frac{1}{2} \lambda_{HX_2} v^2 + \frac{1}{2} \lambda_{\Phi X_2} v^2_X -\mu^2_{X_1} - \frac{1}{2} \lambda_{HX_1} v^2 - \frac{1}{2} \lambda_{\Phi X_1} v^2_X  \right)^2 + 2\mu^2_{X_1 X_2 \Phi} v^2_X }~.  
\label{SDMmass}
\end{split}
\end{equation} 
Therefore, the lighter one $X'_1$ can be a DM candidate and 
we will discuss its stability in the next section.

We then turn to $X'_1$, $X'_2$ interaction terms. Except for the four-point interactions with $X'_1$, $X'_2$ only, there are also $X'_1$, $X'_2$ interactions with the $U(1)_X$ gauge boson and $h_{1,2}$. Because of the DFCNC interaction term, there are not only diagonal interactions $X'_{1,2}\partial^{\mu}X'^*_{1,2}C_{\mu}$, $X'^*_{1,2}X'_{1,2}h_{1,2}$ but also off-diagonal interaction $\left( X'_1 \partial^\mu X'^*_2 + X'_2 \partial^\mu X'^*_1 \right) C_{\mu}$, $\left( X'^*_1 X'_2 + X'^*_2 X'_1 \right) h_{1,2}$ which are different from the usual DM models. 
The complete form of these interactions are shown in Appendix~\ref{Sec:App2}.
It shows a phenomenologically distinguishable feature of this model compared with ordinary dark sectors with $U(1)_X$ gauge symmetry~\cite{Ghorbani:2015baa,  Bell:2016fqf,  Duerr:2016tmh} and inelastic DM models~\cite{Baek:2014kna,  Baek:2020owl,  Kang:2021oes}.  In the former (latter), there is only diagonal (off-diagonal) interactions between DM sectors and the $U(1)_X$ gauge boson. Moreover, both of them only have diagonal interactions between DM sectors and $h_{1,2}$. We will explore effects from these interactions in Sec.~\ref{Sec:constraints} and Sec.~\ref{Sec:BelleII}.

Notice we have other choices for the trilinear $ X_1 $, $ X_2 $ and $ \Phi $ coupling terms : $ X_1 X^{\ast}_2 \Phi $,  
$ X_1 X_2 \Phi^{\ast} $ and $ X_1 X_2 \Phi $ for different $ U(1)_X $ charge assignments of $ \Phi $. 
For $ X_1 X_2 \Phi^{\ast} $ and $ X_1 X_2 \Phi $ operators after the SSB, we will get real scalar DM candidate(s)  instead of a complex scalar DM one 
in Eq.(\ref{eq:potentail1}). 
Since the analysis of these models are similar to present setup,  we will not repeat them again here and only consider a representative one.

Finally, since all SM fermions don't carry $U(1)_X$ charges, the only way for the new gauge boson $ C_\mu $ and SM fermions to interact is via the kinetic mixing between $ B_{\mu\nu} $ and 
$ C_{\mu\nu} $. The Lagrangian density of this part can be represented as
\begin{equation}
{\cal L}_{C,\text{gauge}} = -\frac{1}{4}  C_{\mu\nu}C^{\mu\nu} 
-\frac{\sin\epsilon}{2} B_{\mu\nu}C^{\mu\nu},
\label{eq:Zps}
\end{equation}
where $ \epsilon $ is the kinetic mixing parameter between these two $U(1)$s.
If we apply the linear order approximation in $ \epsilon $ and assume that Z boson is much heavier than the dark photon, the extra interaction terms for SM fermions and the dark 
photon can be written as
\begin{equation}
{\cal L}_{A^{\prime}f\overline{f}} = \epsilon e c_W \sum_{f} x_f \overline{f} \slash{A}^{\prime}f,
\label{eq:Zpffbar}
\end{equation}
where $ c_W $ is the weak mixing angle and $ x_l = -1 $, $ x_{\nu} = 0 $, $ x_q = \frac{2}{3} $ or $ -\frac{1}{3} $ depending on the electrical charge of quark.
The dark photon mass can be approximated as
\begin{equation}
m_{A^{\prime}} \simeq g_X (1-q_X) v_X.
\label{eq:Zpmass}
\end{equation}
For $q_X = 1$, the dark charge of the dark Higgs becomes ``0'' and dark photon becomes 
massless. Therefore another dark Higgs with nonzero dark charge has to be introduced 
if we like to have massive dark photon. 
Notice that the correction from the kinetic mixing term is second order in $ \epsilon $ 
which can be safely neglected here.

\subsection{Fermionic C2CDM models}\label{Sec:FC2CDM} 

\begin{table}[t!]
  \begin{tabular}{l @{\extracolsep{0.2in}} r r r}
   \hline
Fields            & $ \psi_1 $ & $ \psi_2 $ & $ \Phi $ \\ \hline
$ U(1)_X $ charge & $ 1 $   & $ q_X $ & $ 1-q_X $ \\ \hline 
    \end{tabular}
    \caption{\small
    The associated $ U(1)_X $ charges for fields in fermion C2CDM models.}
\label{Tab:U1fcoup}
\end{table}

We define two SM singlet Dirac fermion fields, $ \psi_1 $, $ \psi_2 $ and a SM singlet complex scalar dark Higgs field, $ \Phi $ with $ U(1)_X $ charges in 
Table.~\ref{Tab:U1fcoup}.  We assume $ q_X $ is a real number and $ q_X\neq 0, 1 $ 
such that all $ \psi_1 $, $ \psi_2 $ and $ \Phi $ are charged under $ U(1)_X $. 
Again, we don't consider $ q_X = -1 $ which will induce $ \overline{\psi^c_1}\psi_1  
\Phi^{\ast} $, $ \overline{\psi^c_2}\psi_2\Phi $ operators as in the fermion inelastic DM models~\cite{Baek:2020owl,  Kang:2021oes,  Ko:2019wxq}. Notice all SM fields don't 
carry the $ U(1)_X $ charge in fermion C2CDM models. We can take this model as a dark 
QED with DFCNC interactions for two generation dark fermion fields. 

The scalar part and DM sectors of the renormalizable and gauge invariant Lagrangian density are 
\begin{align} 
{\cal L}_{scalar} = & |D_{\mu}H|^2 + |D_{\mu}\Phi|^2 +\mu^2_H H^{\dagger}H -\lambda_H (H^{\dagger}H)^2 
\nonumber  \\ & 
+\mu^2_{\Phi}\Phi^{\ast}\Phi -\lambda_{\Phi}(\Phi^{\ast}\Phi)^2 -\lambda_{H\Phi}(H^{\dagger}H)(\Phi^{\ast}\Phi)
\label{eq:Lf}
\end{align} 
and 
\begin{equation}
{\cal L}_{\psi} = \overline{\psi_1}(i\slash{D}-M_{\psi_1})\psi_1 +\overline{\psi_2}(i\slash{D}-M_{\psi_2})\psi_2 -(f\overline{\psi_1}\psi_2\Phi +H.c.) 
\label{eq:LDM}
\end{equation} 
with 
\begin{align} 
& D_{\mu}\Phi = (\partial_{\mu}+ig_X (1-q_X) C_{\mu})\Phi,
\nonumber  \\ &
D_{\mu}\psi_1 = (\partial_{\mu}+ig_X C_{\mu})\psi_1,
\nonumber  \\ & 
D_{\mu}\psi_2 = (\partial_{\mu}+ig_X q_X C_{\mu})\psi_2.
\label{eq:covariant2}
\end{align}
Again, if we change the $ U(1)_X $ charge assignment in Table~\ref{Tab:U1fcoup} such that the last term in Eq.(\ref{eq:LDM}) is not allowed, this model turns to the simple $ U(1)_1\times U(1)_2 $ two-component fermion DM model.

We expand $H$ and $\Phi$ fields around the vacuum with the unitary gauge as shown in Eq.(\ref{eq:expand}), and the mass term of DM sectors can be written as 
\begin{equation}
{\cal L}_{\text{DM,mass}} = - 
\left( \overline{\psi_1} \overline{\psi_2} \right)   
\left(
\begin{tabular}{cc}
$M_{\psi_1}$ & $f v_X /\sqrt{2}$ 
\\
$f v_X /\sqrt{2}$ & $M_{\psi_2}$ 
\end{tabular}
\right) 
\left(
\begin{tabular}{c}
$\psi_1$
\\
$\psi_2$
\end{tabular}
\right)
\label{eq:mass1}
\end{equation}
and define the mass eigenstates $\psi^\prime_1$ and $\psi^\prime_2$ from the interaction eigenstates $\psi_1$ and $\psi_2$ as
\begin{equation}
\psi^\prime_1 = \psi_1 \cos \theta_\psi - \psi_2 \sin \theta_\psi , \quad \psi^\prime_2 =  \psi_1 \sin \theta_\psi + \psi_2 \cos \theta_\psi~,
\end{equation}
where the mixing angle satisfies 
\[
\tan 2\theta_\psi = \frac{\sqrt{2} f v_X}{M_{\psi_2} - M_{\psi_1}} .
\] 
Their mass eigenvalues can be solved as
\begin{equation}
M_{\psi^\prime_1,\psi^\prime_2} = \frac{1}{2} \left(  M_{\psi_1}+M_{\psi_2} \mp \sqrt{\left( M_{\psi_1}-M_{\psi_2}  \right)^2+ 2f^2 v^2_X} \right)~. 
\label{FDMmass}
\end{equation} 
The lighter one $\psi^\prime_1$ can be a DM candidate, and its stability will be 
discussed in the following section.
Similar to the situation of scalar C2CDM models in Sec.~\ref{Sec:SC2CDM},  there are both diagonal and off-diagonal $\psi^\prime_1$, $\psi^\prime_2$ interactions with the $A'$ and $h_{1,2}$. The complete form of these interactions are shown in Appendix~\ref{Sec:App2}.

Again, we have other choices for the Yukawa coupling term of $ \psi_1 $, $ \psi_2 $ and $ \Phi $ : $ \overline{\psi_1}\psi_2 \Phi^{\ast} $, $ \overline{\psi^c_1}\psi_2 \Phi $ and $ \overline{\psi_1^c}\psi_2 \Phi^{\ast} $ for different $ U(1)_X $ charge assignments of $ \Phi $. 
For $ \overline{\psi^c_1}\psi_2 \Phi $ and $ \overline{\psi_1^c}\psi_2 \Phi^{\ast} $ Yukawa coupling terms after the SSB, we will receive Majorana fermion DM candidate(s) instead of a Dirac fermion DM one in Eq.~(\ref{FDMmass}).
Finally, the dark photon mass and its interactions with SM fermions are the same as Eqs.~(\ref{eq:Zpmass}) and~(\ref{eq:Zpffbar}), respectively.

\section{The stability of DM candidates, DM relic density and direct detection constraints}\label{Sec:stability}

After the SSB of $ U(1)_X $ gauge symmetry, we expect the accidentally residual global $ U(1) $ symmetry for $X^\prime_{1,2}$ in scalar C2CDM and $\psi^\prime_{1,2}$ in fermion C2CDM,
\begin{equation}
X^\prime_{1,2} \rightarrow e^{i \phi_X} X^\prime_{1,2},\quad 
\psi^\prime_{1,2} \rightarrow e^{i \phi_{\psi}} \psi^\prime_{1,2}
\end{equation} 
such that $ X^{\prime}_1 $ or $ \psi^{\prime}_1 $ can be the DM candidate\footnote{For the model with $ X_1 X_2 \Phi $ operator (scalar DM) or $ \overline{\psi^c_1}\psi_2 \Phi $ operator (fermion DM), we expect the accidentally residual $ Z_2 $ symmetry instead of global $ U(1) $ symmetry after the SSB of $ U(1)_X $ gauge symmetry.}.  
These  accidental $U(1)$ symmetries are defined in the mass basis 
\footnote{DM fields with prime ' are in the mass basis, and those without prime are 
in the interaction basis.}, and they appear because DM fields are complex 
scalar or Dirac fields so that global charges are well defined.
However, for the specific value of $ q_X $, it is possible that this  global $ U(1) $ symmetry is broken by dimension-three (dim-3) and dimension-five (dim-5) operators, which 
would make DM candidate decay too fast unless its mass is very tiny \cite{Baek:2013qwa}. 
Therefore, we first discuss the stability of DM candidates in C2CDM models and then calculate DM relic density and comment on DM direct detection in this section.

For scalar C2CDM models in Sec.~\ref{Sec:SC2CDM}, possible dim-3 operators to break the global $ U(1) $ symmetry are 
\begin{align} 
& q_X = \frac{3}{2} : \quad \mu_{X_1\Phi\Phi}X_1 (\Phi)^2 +H.c., 
\nonumber  \\ & 
q_X = \frac{1}{2} : \quad \mu^{\prime}_{X_1\Phi^{\ast}\Phi^{\ast}}X_1 (\Phi^{\ast})^2 +H.c., 
\nonumber  \\ & 
q_X = 2 : \quad \mu_{X_2\Phi\Phi}X_2 (\Phi)^2 +H.c.,
\nonumber  \\ & 
q_X = \frac{2}{3} : \quad \mu^{\prime}_{X_2\Phi^{\ast}\Phi^{\ast}}X_2 (\Phi^{\ast})^2 +H.c.. 
\label{Sdanger3}
\end{align}
Also there are possible dim-5 operators to break the global $ U(1) $ symmetry, 
\begin{align} 
& q_X = \frac{3}{2} : \quad \frac{c_{s1}}{\Lambda}X_1 (\Phi)^2\left[ (H^{\dagger}H)+(\Phi^{\ast}\Phi)\right] +H.c., 
\nonumber  \\ & 
q_X = \frac{1}{2} : \quad \frac{c^{\prime}_{s1}}{\Lambda}X_1 (\Phi^{\ast})^2\left[ (H^{\dagger}H)+(\Phi^{\ast}\Phi)\right] +H.c., 
\nonumber  \\ & 
q_X = 2 : \quad \frac{c_{s2}}{\Lambda}X_2 (\Phi)^2\left[ (H^{\dagger}H)+(\Phi^{\ast}\Phi)\right] +H.c.,
\nonumber  \\ & 
q_X = \frac{2}{3} : \quad \frac{c^{\prime}_{s2}}{\Lambda}X_2 (\Phi^{\ast})^2\left[ (H^{\dagger}H)+(\Phi^{\ast}\Phi)\right] +H.c., 
\label{Sdanger5}
\end{align}
where $ \Lambda $ is the cut-off scale that is presumably bounded from above by
Planck scale. We list possible $ X^{\prime}_1 $ decay channels and partial decay widths in the Appendix~\ref{Sec:App2}.

For fermionic C2CDM models in Sec.~\ref{Sec:FC2CDM}, possible dim-5 operators to break the global $ U(1) $ symmetry are 
\begin{align} 
& q_X = 2 : \quad \frac{c_{f1}}{\Lambda}(\overline{L}\tilde{H})(\psi_{1R}\Phi) +H.c., 
\nonumber  \\ & 
q_X = \frac{1}{2} : \quad \frac{c_{f2}}{\Lambda}(\overline{L}\tilde{H})(\psi_{2R}\Phi^{\ast}) +H.c., 
\label{Fdanger5}
\end{align}
where $ \psi_{1,2R} = \frac{1+\gamma_5}{2}\psi_{1,2} $. Again, possible $ \psi^{\prime}_1 $ decay channels and partial decay widths are listed in the Appendix~\ref{Sec:App2}\footnote{
If $ M_{\psi^{\prime}_1} < m_e $, $ \psi_{1,2} $ can mix with SM neutrinos such that $ \psi^{\prime}_{1,2}\rightarrow 3\nu $ and $ \psi^{\prime}_{1,2}\rightarrow\nu\gamma\gamma $ are possible. However, for $\psi^{\prime}_1$ with ${\cal O}(10)$ keV, if its lifetime is much longer than the age of universe, $ \psi^{\prime}_1 $ can still become a good DM candidate as the light sterile neutrino DM~\cite{Drewes:2016upu,Boyarsky:2018tvu}. 
}.

In order to avoid dangerous operators in Eqs. (\ref{Sdanger3}),~(\ref{Sdanger5}) and~(\ref{Fdanger5}) which break the global $U(1)$ symmetry,  we assign $ q_X = -4 $ for both scalar and fermion C2CDM models as examples. 
We then turn to the issue of DM relic density.  
Depending on the the mass scale of $X^{\prime}_{1,2}$ ($\psi^{\prime}_{1,2}$) and their mass spectrum with $A'$ and $h_{1,2}$ in scalar (fermion) C2CDM models, there are plenty of possible DM annihilation mechanisms for the freeze-out scenario : 
\begin{itemize}
\item $ M_{X^{\prime}_1, \psi^{\prime}_1} < m_{A^{\prime}}, m_{h_{1,2}} $ and $ \Delta_{X,\psi}\gg 0.5 M_{X^{\prime}_1, \psi^{\prime}_1}$ \\
In this case, C2CDM models will reduce to two portals model (Higgs portal and vector portal)~\cite{Ghorbani:2015baa,Bell:2016fqf,  Duerr:2016tmh} and $X^{\prime}_2, \psi^{\prime}_2$ will not play the role for the relic density.  
The dominant annihilation channels are $ X^{\prime}_{1}X^{\prime\ast}_{1} (\psi^{\prime}_{1}\overline{\psi^{\prime}}_{1})\rightarrow f\overline{f} $ via s-channel $ A^{\prime}, h_{1,2} $. 

\item $ M_{X^{\prime}_1, \psi^{\prime}_1} < m_{A^{\prime}}, m_{h_{1,2}} $ and $ \Delta_{X,\psi}\lesssim 0.5 M_{X^{\prime}_1, \psi^{\prime}_1}$ \\
Both $X^{\prime}_1$, $X^{\prime}_2$ (or $\psi^{\prime}_1$, $\psi^{\prime}_2$) can contribute to the DM relic density.  
The major annihilation and co-annihilation channels are $ X^{\prime}_{1}X^{\prime\ast}_{1}( \psi^{\prime}_{1}\overline{\psi^{\prime}}_{1})\rightarrow f\overline{f} $ and $ X^{\prime}_1 X^{\prime\ast}_2 (\psi^{\prime}_1 \overline{\psi^{\prime}}_2)\rightarrow f\overline{f} $  via s-channel $ A^{\prime}, h_{1,2} $.
 
\item $ M_{X^{\prime}_1, \psi^{\prime}_1} > m_{A^{\prime}}, m_{h_{1,2}} $ and $ \Delta_{X,\psi}\gg 0.5 M_{X^{\prime}_1, \psi^{\prime}_1}$ \\ 
Again, $X^{\prime}_2, \psi^{\prime}_2$ will not play the role for the relic density. The major annihilation channels can change to $ X^{\prime}_{1}X^{\prime\ast}_{1} (\psi^{\prime}_{1}\overline{\psi^{\prime}}_{1})\rightarrow A^{\prime}A^{\prime} $,  $ X^{\prime}_{1}X^{\prime\ast}_{1} (\psi^{\prime}_{1}\overline{\psi^{\prime}}_{1})\rightarrow h_{1,2} h_{1,2} $ and $ X^{\prime}_{1}X^{\prime\ast}_{1} (\psi^{\prime}_{1}\overline{\psi^{\prime}}_{1})\rightarrow h_2 A^{\prime} $~\cite{Pospelov:2007mp,  Bell:2016fqf,  Baek:2020owl,  Duerr:2020muu}. 

\item $ M_{X^{\prime}_1, \psi^{\prime}_1} > m_{A^{\prime}}, m_{h_{1,2}} $ and $ \Delta_{X,\psi}\lesssim 0.5 M_{X^{\prime}_1, \psi^{\prime}_1}$ \\ 
Apart from the previous annihilation channels, an extra new co-annihilation channel $ X^{\prime}_{1}X^{\prime\ast}_{2} (\psi^{\prime}_{1}\overline{\psi^{\prime}}_{2})\rightarrow h_{1,2} A^{\prime} $  is also possible. 

\item $ 1\lesssim m_{A'}/M_{X^{\prime}_1, \psi^{\prime}_1}\lesssim 2$ and/or $ \frac{1}{2}\lesssim m_{A'}/\left(M_{X^{\prime}_1, \psi^{\prime}_1} + M_{X^{\prime}_2, \psi^{\prime}_2}\right) \lesssim 1$ with $ \Delta_{X,\psi}\ll M_{X^{\prime}_1, \psi^{\prime}_1} $ \\ 
For the sub-GeV DM, there are also novel $3\rightarrow 2$ number-changing processes~\cite{Cline:2017tka,Fitzpatrick:2020vba,Fitzpatrick:2021cij} and forbidden channels~\cite{DAgnolo:2015ujb} in C2CDM models.  

\item $ \Delta_{X,\psi}\rightarrow 0 $ \\ 
Both $ X^{\prime}_1 (\psi^{\prime}_1) $ and $ X^{\prime}_2 (\psi^{\prime}_2) $ can be DM candidates and form the two-component DM scenario in this situation.  
\end{itemize}

In this work, we will first focus on the novel collider signatures of $A'$, $h_2$ semi-visible decay channels for $m_{A',h_2} > 2 M_{X^{\prime}_2, \psi^{\prime}_2}$. 
Therefore, only $ X^{\prime}_{1, 2}X^{\prime\ast}_{1, 2}\rightarrow f\overline{f} $ ($ \psi^{\prime}_{1, 2}\overline{\psi^{\prime}}_{1, 2}\rightarrow f\overline{f} $) and $ X^{\prime}_1 X^{\prime\ast}_2\rightarrow f\overline{f} $ ($ \psi^{\prime}_1 \overline{\psi^{\prime}}_2\rightarrow f\overline{f} $) via s-channel $ A^{\prime}, h_{1,2} $ are relevant for our DM relic density calculation here. 
However, we will explore other DM annihilation mechanisms in Sec.~\ref{Sec:511X1T} motivated by the explanation of $511$ keV $\gamma$-ray line and XENON1T excess in C2CDM models.

In order to avoid severe DM direct detection constraints, we choose $ \sin\theta_{X,\psi} = 1/\sqrt{1-q_{X,\psi}} $ as a benchmark point such that $ X'_1 X'^{\ast}_1 A' $, 
$ \psi'_1 \overline{\psi'_1} A' $ couplings in the tree-level are equal to zero. 
This choice is also important for the DM mass less than $10$ GeV in the fermionic C2CDM 
models since the s-wave $ \psi'_1\overline{\psi'_1}\rightarrow A'\rightarrow f\overline{f} $ annihilation process has already been ruled out by CMB constraint~\cite{Ade:2015xua}.  For scalar C2CDM models, even though $ X'_1 X'^{\ast}_1\rightarrow A'\rightarrow f\overline{f} $ annihilation process is p-wave, $ M_{X'_1}\lesssim 10 $ MeV is not preferred from BBN constraint~\cite{Depta:2019lbe,  Krnjaic:2019dzc}. 
In order to study the allowed parameter space of scalar C2CDM models, we fix the parameters as following:
\begin{equation}
\begin{split}
& m_{h_1} = 125 \, {\rm GeV}, \ m_{h_2} = 3M_{X^\prime_1}, \ \sin \theta_h = 10^{-3}, \ \alpha_X = 0.1,
\\ &  \lambda_{X_1} = \lambda_{X_2} = \lambda_{X_1 X_2} = \lambda_{HX_1} = \lambda_{\Phi X_1} = \lambda_{HX_2} = \lambda_{\Phi X_2} = 0,
\end{split}
\end{equation} 
to simplify our analysis. 
We vary $M_{X^\prime_1}$ from $0.1$ to $30 \, {\rm GeV}$, $\epsilon$ from $10^{-4}$ to $0.1$, $\Delta_{X}/M_{X^\prime_1}$ from $0.01$ to $1$ and $m_{A^\prime}$ from $0.3$ to $90 \, {\rm GeV}$.
The effects from non-vanishing quartic couplings, $ \lambda_{X_1}$, 
$\lambda_{X_2}$, $\lambda_{X_1 X_2}$, $\lambda_{HX_1}$, $\lambda_{\Phi X_1}$, 
$\lambda_{HX_2}$ and $\lambda_{\Phi X_2}$ are discussed in the Appendix~\ref{Sec:App3}.
Similarly, we fix parameters in the fermionic C2CDM models as following:
\begin{equation}
m_{h_1} = 125 \, {\rm GeV}, \ m_{h_2} = 3M_{\psi^\prime_1}, \ \sin \theta_h = 10^{-3}, \ \alpha_X = 0.1.
\end{equation}
We vary $M_{\psi^\prime_1}$ from $0.1$ to $30 \, {\rm GeV}$, $\epsilon$ from $10^{-4}$ to $0.1$, $\Delta_{\psi}/M_{\psi^\prime_1}$ from $0.01$ to $1$ and $M_{A^\prime}$ from $0.3$ to $90 \, {\rm GeV}$\footnote{In order for the comparison, an arbitrary $\sin\theta_{X,\psi}$ is also studied. However, we find the DM annihilation mainly comes from $X'_1 X'^{\ast}_1$ or $\psi'_1 \overline{\psi'_1}$ and the contributions from DM co-annihilation are much suppressed.}.

Furthermore, we classify the parameter space according to the size of $ \Delta_{X,\psi} $. For $\Delta_{X,\psi}\gtrsim 0.5 M_{X^{\prime}_1, \psi^{\prime}_1}$, there are only $ X^{\prime}_{1}X^{\prime\ast}_{1} $ (or $ \psi^{\prime}_{1}\overline{\psi^{\prime}_{1}} $) annihilation. 
For $2 m_e < \Delta_{X,\psi}\lesssim 0.5 M_{X^{\prime}_1, \psi^{\prime}_1}$, the $ X^{\prime}_{1}X^{\prime\ast}_{2} $ (or $ \psi^{\prime}_{1}\overline{\psi^{\prime}_{2}} $) co-annihilation channel becomes the dominant one.  
For the assumption $ \sin\theta_{X,\psi} = 1/\sqrt{1-q_{X,\psi}} $,  there is no $ X'_1 X'^{\ast}_1 A' $, $ \psi'_1 \overline{\psi'_1} A' $ couplings at 
tree-level.  The elastic $ X^{\prime}_1$,  $\psi^{\prime}_1 $ and nucleons scattering via $h_2$ is much suppressed because of the small $\sin\theta_{X,\psi}$ and Yukawa couplings.  
On the other hand,  the elastic $ X^{\prime}_1$,  $\psi^{\prime}_1 $ and nucleons scattering can only be generated via one-loop diagrams with $ A' $, whose cross sections are proportional to $ \alpha^2_{\text{em}}\alpha^2_X \epsilon^4 $.  As pointed out in Refs.~\cite{Izaguirre:2015zva, Sanderson:2018lmj,  Berlin:2018jbm,  Duerr:2019dmv},  DM direct detection constraints from this kind of loop processes are still much weaker than the LEP one~\cite{Hook:2010tw} on $ \epsilon $ for $ \alpha_X \sim {\cal O} (1) $.
Hence,  constraints from DM direct detection can be safely ignored in this case as inelastic DM models.  

\begin{figure}
\centering
\includegraphics[width=2.8in]{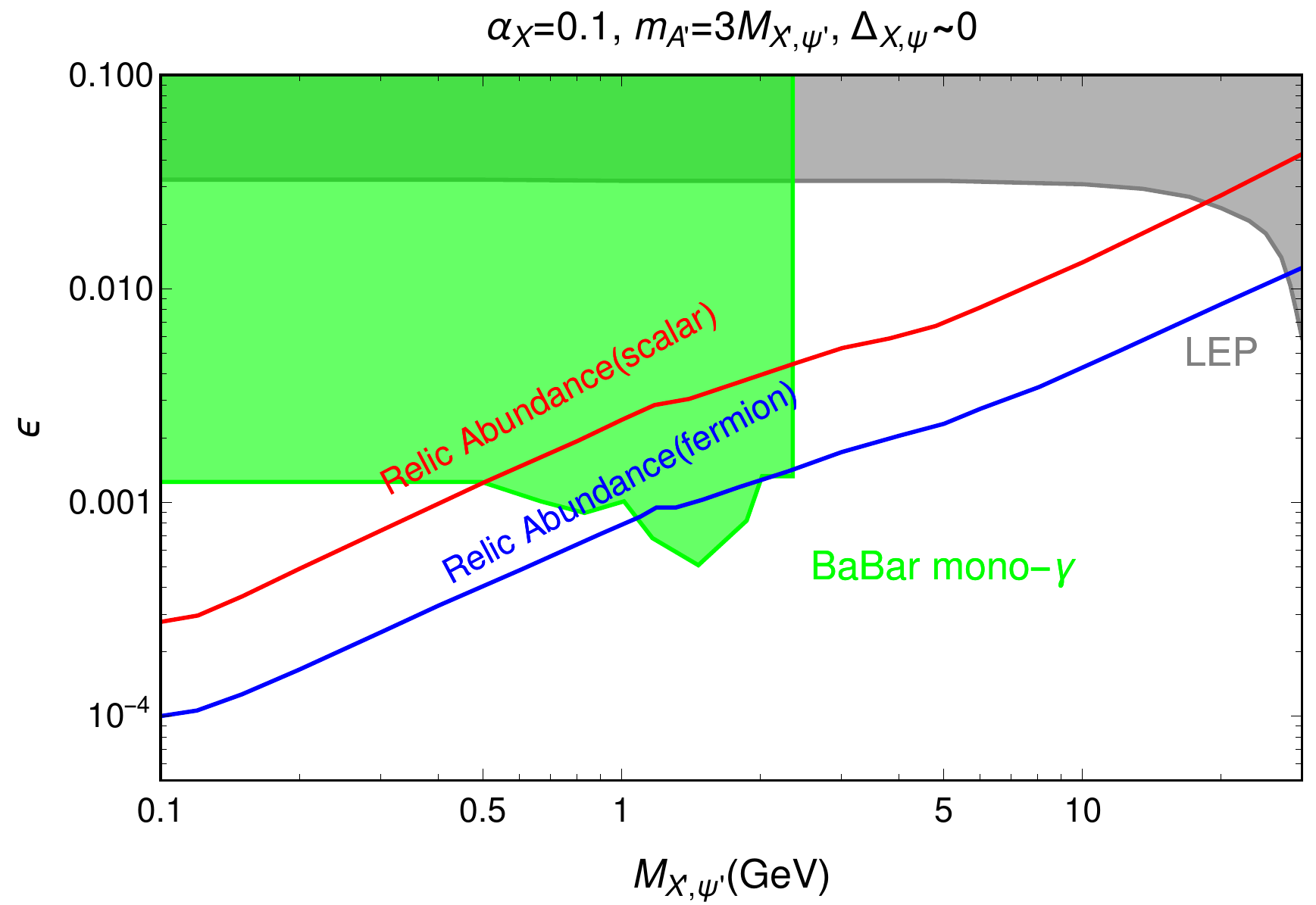} 
\includegraphics[width=3.6in]{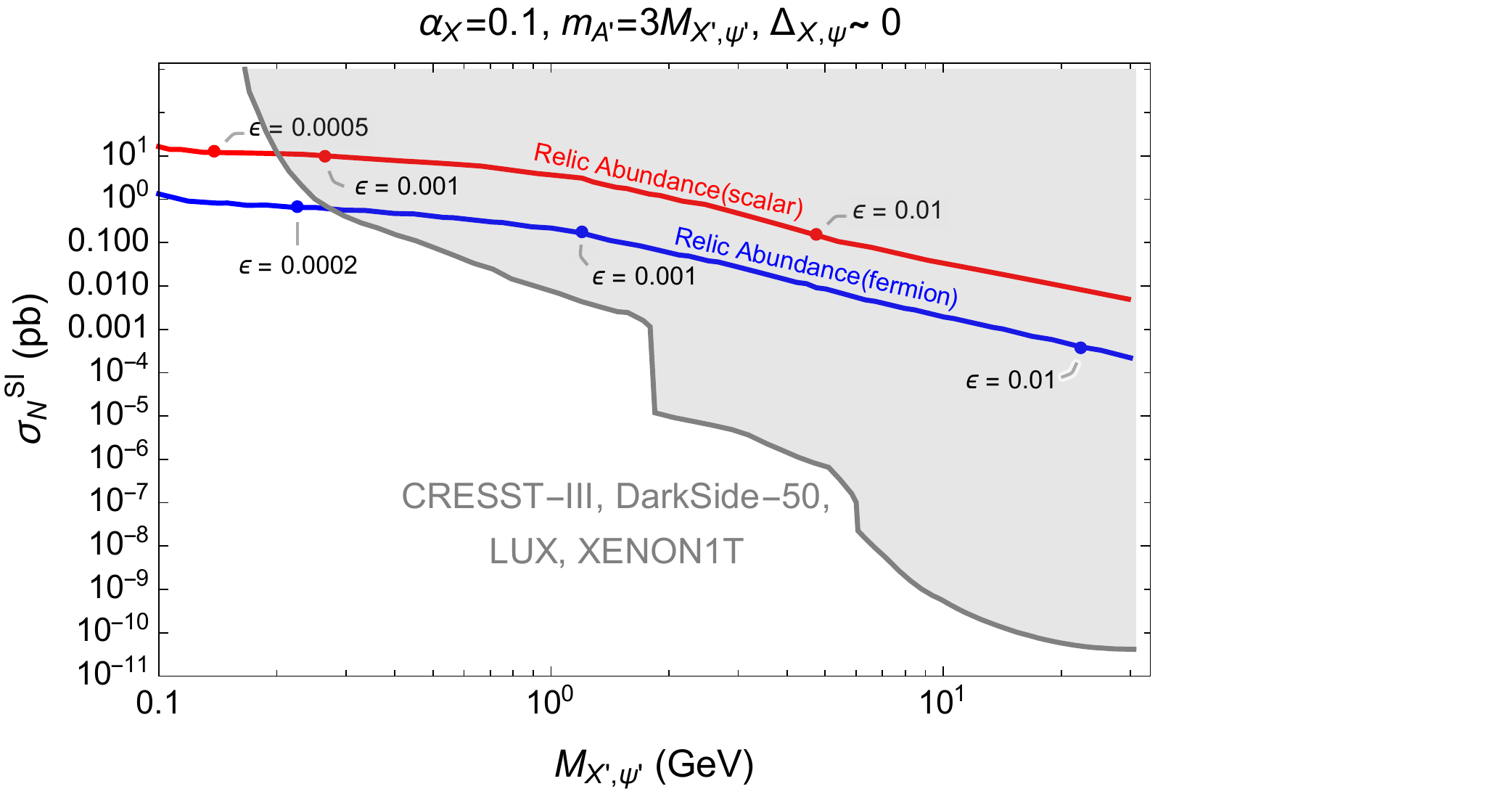}
\caption{ 
The allowed parameter space of $(M_{X',\psi'},\epsilon)$ (left) and $(M_{X',\psi'},\sigma^{\text{SI}}_N)$ (right) with $\alpha_X = 0.1$, $m_{A'} = 3 M_{X',\psi'}$, $\Delta_{X,\psi}\sim 0$. In the left panel, the model-independent LEP bound~\cite{Hook:2010tw}, BaBar mono-$\gamma$ bound~\cite{Lees:2017lec} and observed DM relic abundance lines are included. In the right panel, the constraints from DM direct detection, CRESST-III~\cite{CRESST:2019jnq}, DarkSide-50~\cite{DarkSide:2018ppu}, LUX~\cite{LUX:2018akb}, and XENON1T~\cite{XENON:2018voc} are considered. 
}\label{fig:2DM_con}
\end{figure}

Finally, for $\Delta_{X,\psi}\rightarrow 0$, there is almost no mixing between $X_1, X_2$ (or $\psi_1,\psi_2$). Therefore, our results are close to ordinary two-component DM models and cannot be avoided from strong constraints of DM direct detection.  Here we set $ M_{X_1,\psi_1}\sim M_{X_2,\psi_2} $ ($ \Delta_{X,\psi}\sim 0 $) and $ m_{A'} = 3M_{X'_1,\psi'_1} $ with varying $ \epsilon $.   
The left panel of Fig.~\ref{fig:2DM_con} shows the parameter space $ (M_{X',\psi'}, \epsilon) $ for the observed DM relic density ($ \Omega h^2 = 0.12 $)  
for both scalar and fermion DM cases, with the bounds from the dark photon invisible decay searches 
from LEP~\cite{Hook:2010tw} and BaBar~\cite{Lees:2017lec}. 
In the right panel of Fig.~\ref{fig:2DM_con}, we show the parameter space 
in the $ (M_{X',  \psi'}, \sigma^{\text{SI}}_N) $ plane that is allowed by the observed DM relic density with suitably chosen 
$ \epsilon $ values marked on the curves and direct detection bounds.  In the DM mass region shown in Fig.~\ref{fig:2DM_con}, the relevant DM direct detection constraints are given by CRESST-III~\cite{CRESST:2019jnq} from 0.2 to 2 GeV, DarkSide-50~\cite{DarkSide:2018ppu} from 2 to 5 GeV, LUX~\cite{LUX:2018akb} from 5 to 6 GeV, and XENON1T~\cite{XENON:2018voc} from 6 to 30 GeV.
We notice that DM direct detection constraints are much stronger than collider ones in two-component DM scenario, and $ M_{X'}\gtrsim 0.2 $ GeV and $ M_{\psi'}\gtrsim 0.3 $ GeV have already been ruled out. Therefore, there is no allowed parameter space for the $s$-wave annihilation in case of two-component DM scenario
of fermionic C2CDM models.

\begin{figure}
\centering
\includegraphics[width=2.1in]{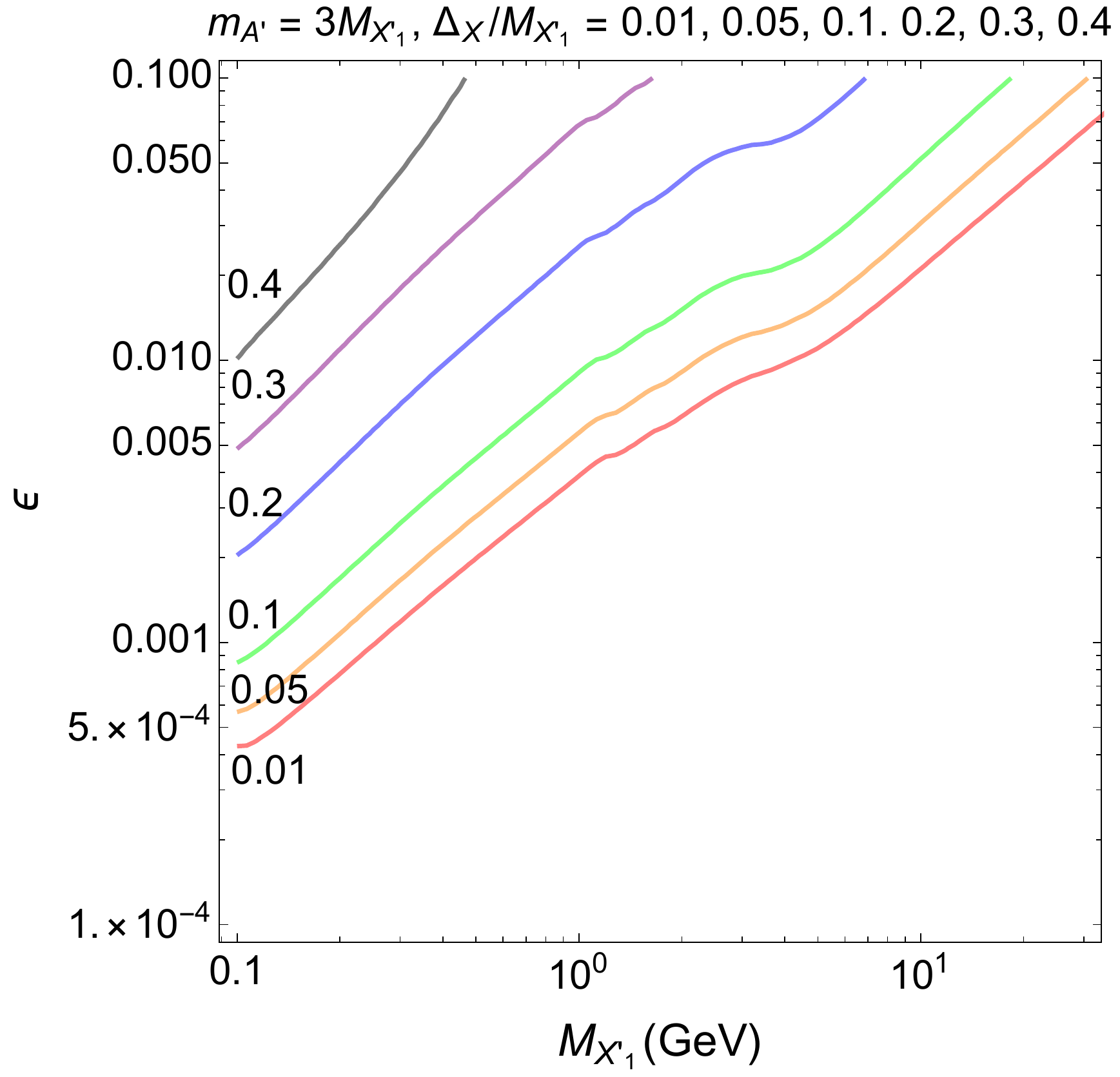} 
\includegraphics[width=2.0in]{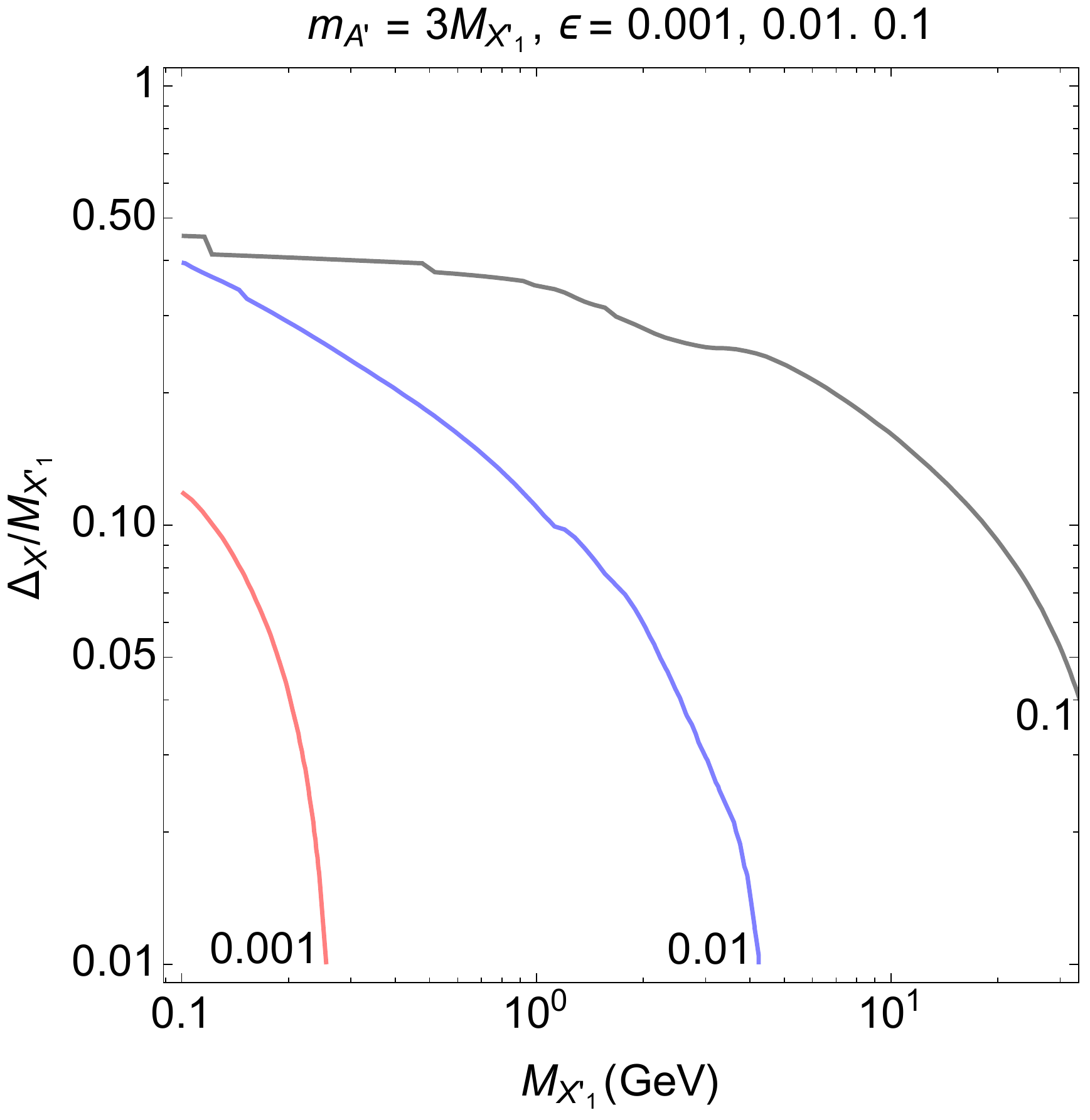}
\includegraphics[width=2.1in]{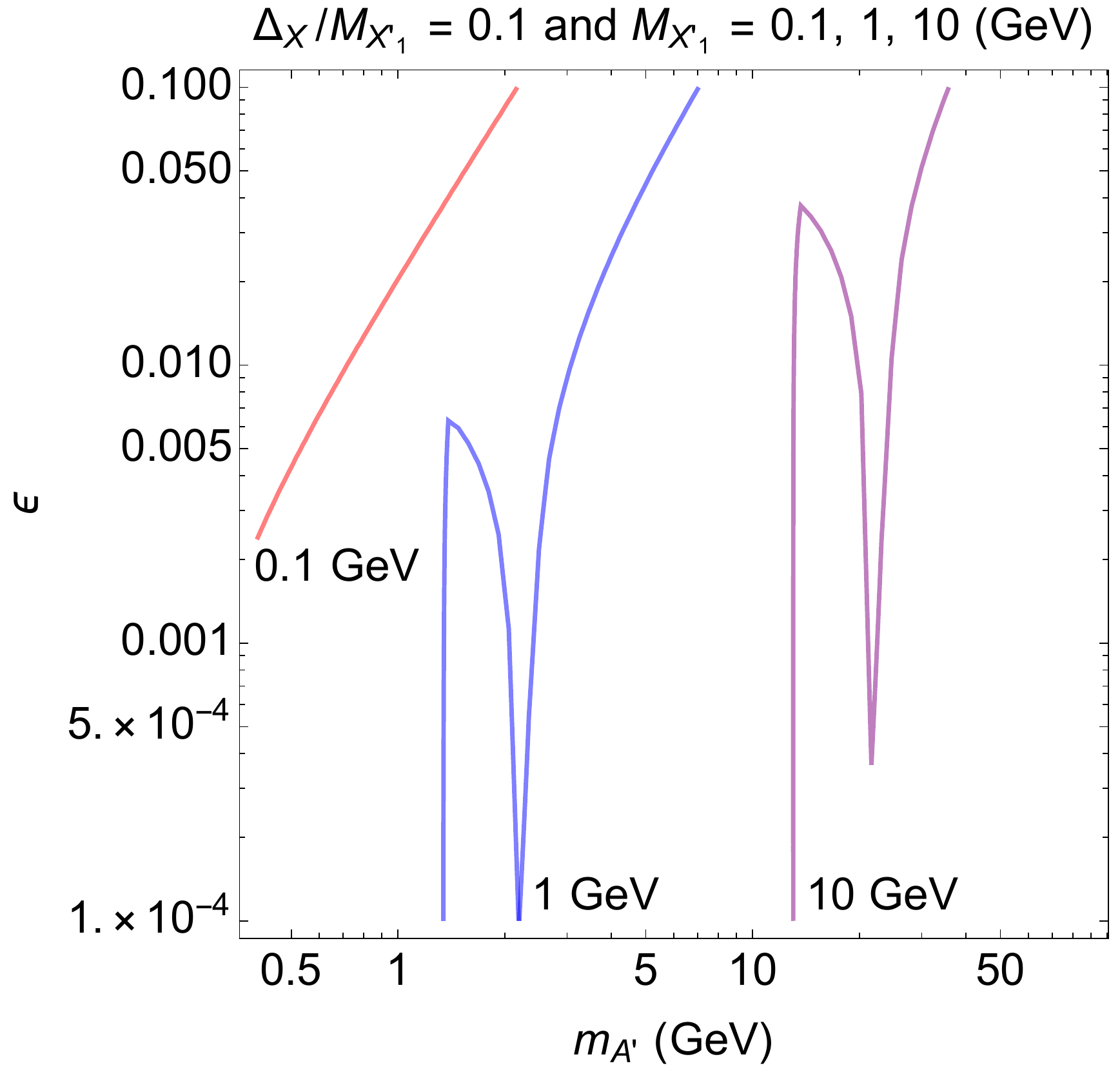}
\includegraphics[width=2.1in]{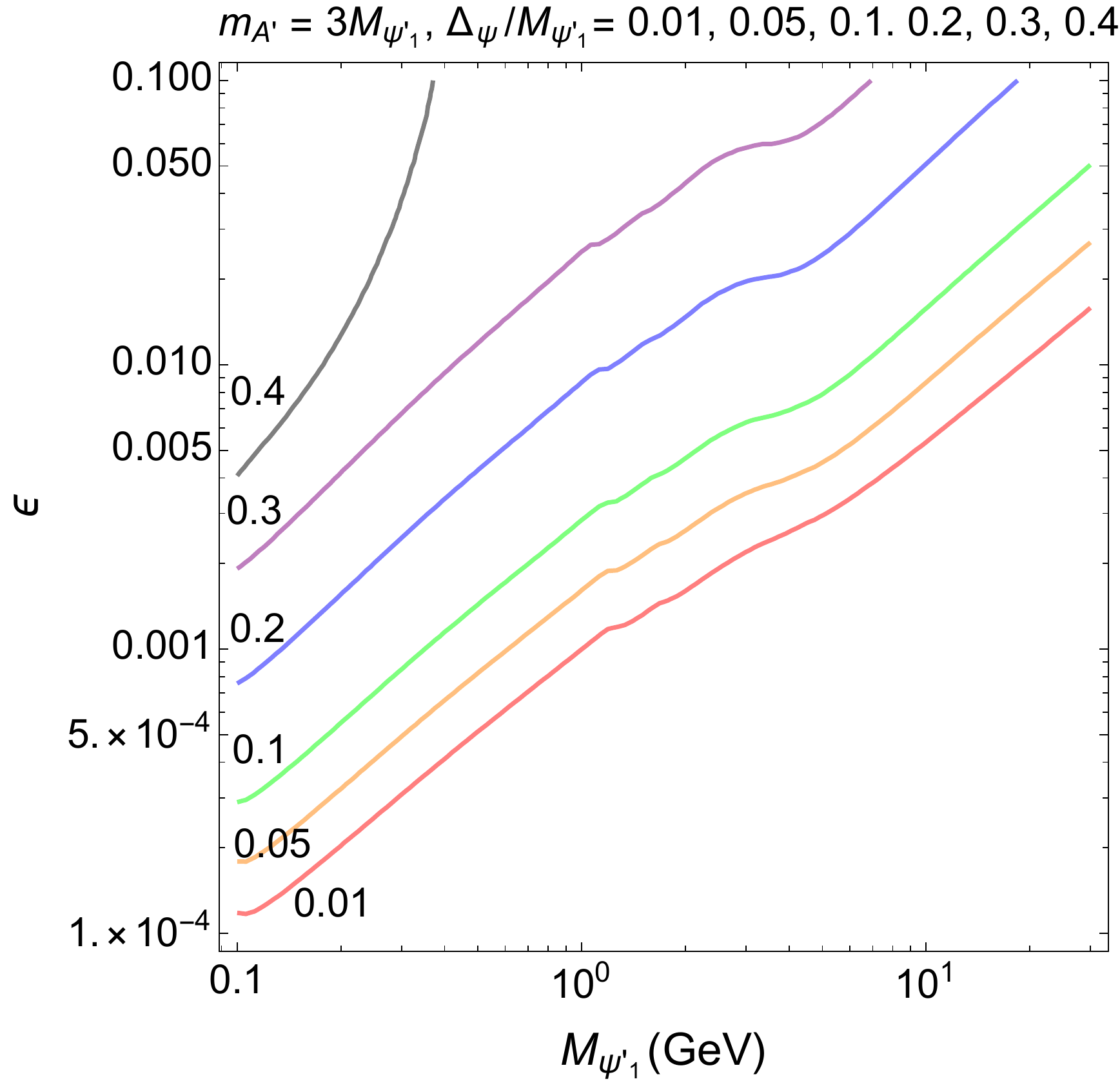}
\includegraphics[width=2.0in]{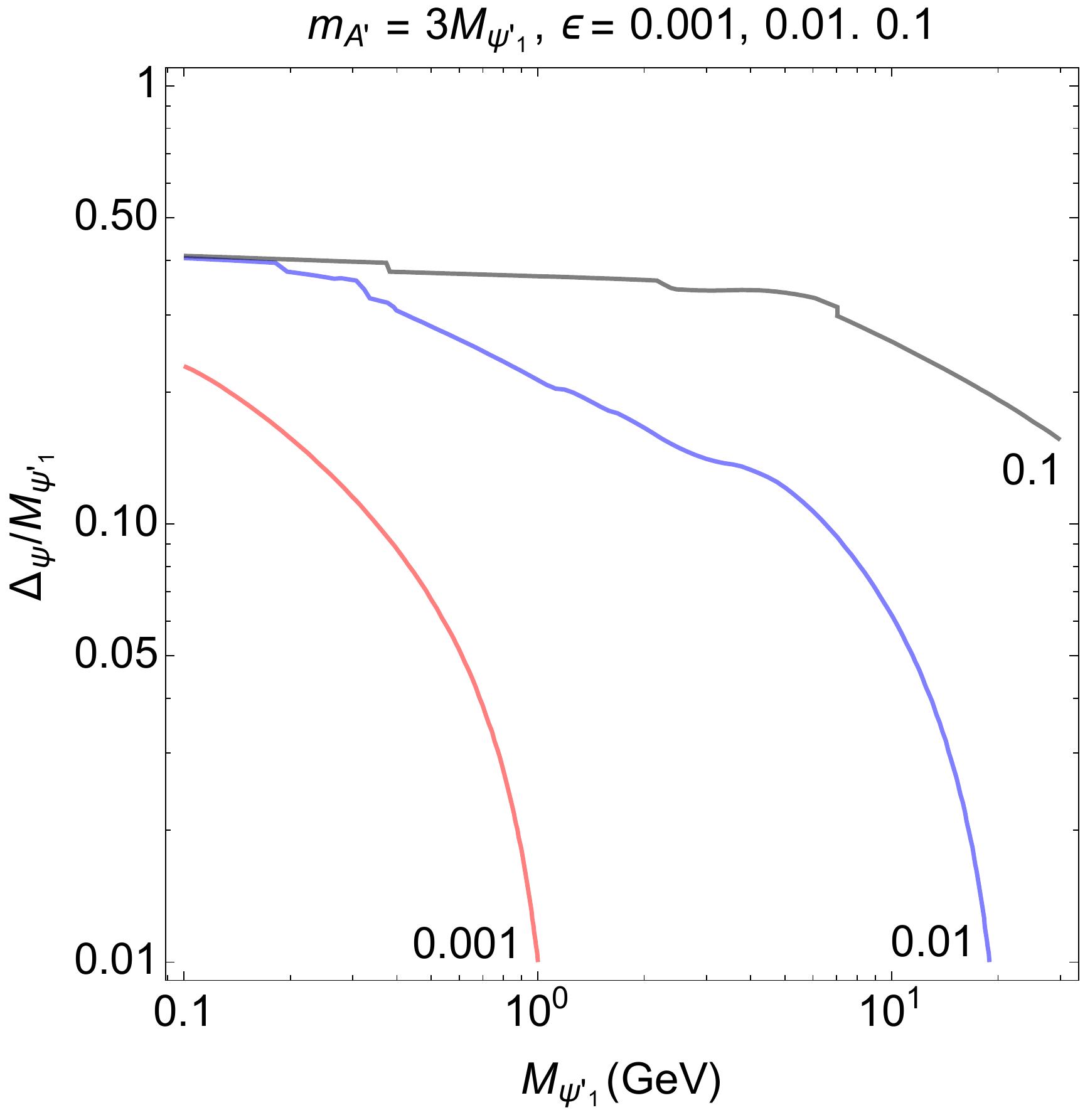}
\includegraphics[width=2.1in]{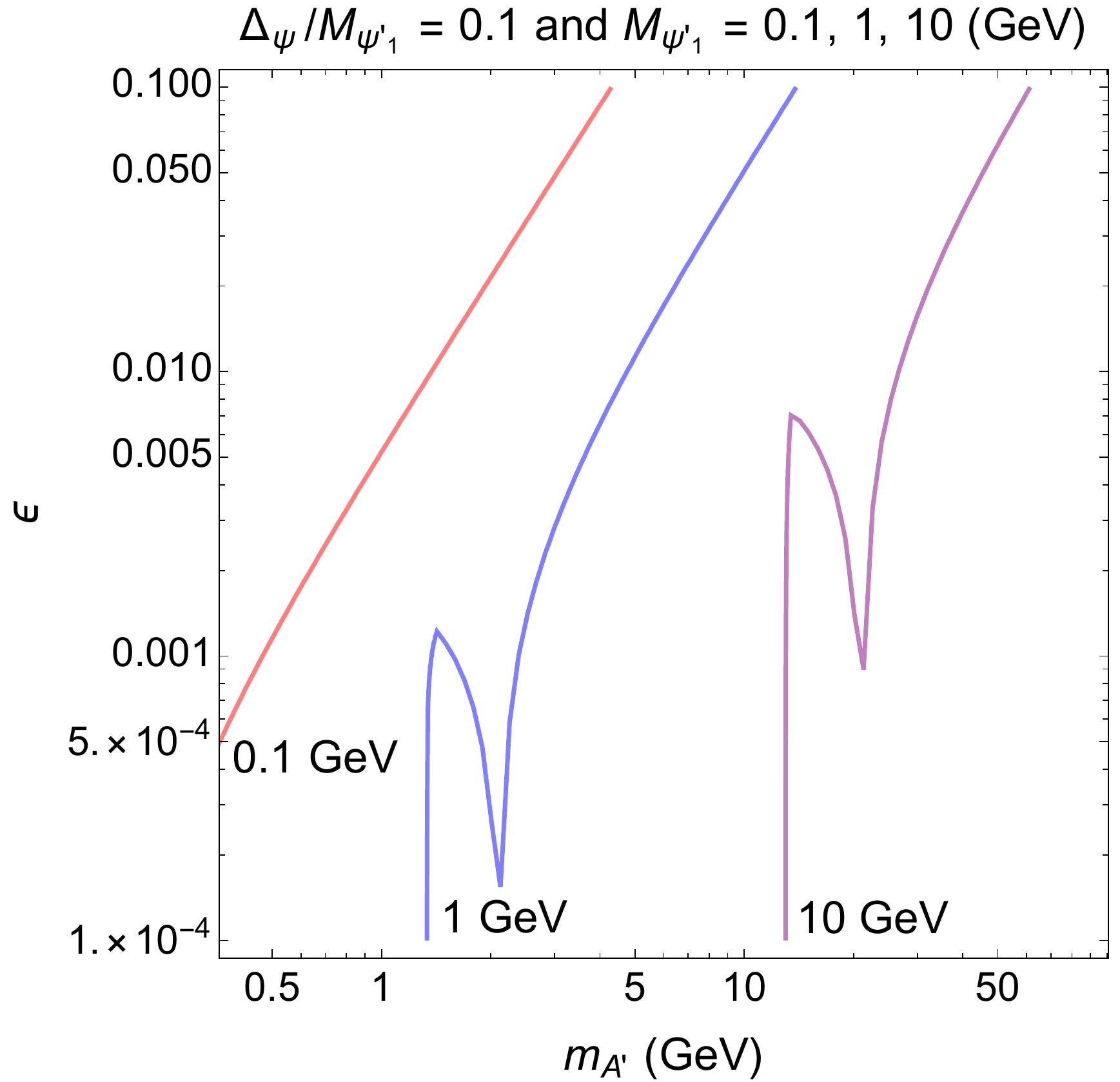}
\caption{ 
Various parameter space in the scalar (upper) and fermionic (lower) C2CDM models. In the left panel, $(M_{X'_1,\psi'_1},\epsilon)$ with $m_{A'} = 3 M_{X'_1,\psi'_1}$ and varying $\Delta_{X,\psi}/M_{X'_1,\psi'_1}$. In the middle panel, $(M_{X'_1,\psi'_1},\Delta_{X,\psi}/M_{X'_1,\psi'_1})$ with $m_{A'} = 3 M_{X'_1,\psi'_1}$ and varying $\epsilon$. In the right panel, $(m_{A'},\epsilon)$ with $\Delta_{X,\psi} = 0.1$ and varying $M_{X'_1,\psi'_1}$.
}\label{fig:relic}
\end{figure}

We then turn to the single-component DM scenario with $\Delta_{X, \psi}\geqslant 2 m_e$. 
In the left and middle panels of Fig.~\ref{fig:relic}, we first fix the relation $m_{A^\prime}=3M_{X^\prime_1,\psi^\prime_1}$ and show the lines which can satisfy $ \Omega_{X'_1, \psi'_1}h^2 = 0.12 $. 
For the parameter space $(M_{X'_1, \psi'_1}, \epsilon)$ in the left panels, we vary 
$ \Delta_{X, \psi} / M_{X'_1, \psi'_1} = 0.01, 0.05, 0.1, 0.2, 0.3, 0.4 $. Once we fix the DM mass, it's clear that the smaller $ \Delta_{X, \psi} $ corresponds to the smaller $ \epsilon $ which is important for searching the displaced vertex signatures at collider and fixed target experiments. Compared with scalar and fermionic C2CDM models, the scalar and fermion pair annihilation cross sections can be scaled by $ \beta^{3/2} $ and $ \beta^{1/2} $ respectively, where $ \beta $ is the velocity of the final state particle in the center-of-momentum frame. Because of this extra $ \beta $ factor for the scalar case, cross sections for  $ X'_1 X'^{\ast}_2\rightarrow f\overline{f} $ are suppressed compared with 
$ \psi'_1 \overline{\psi'_2}\rightarrow f\overline{f} $.
For the parameter space $(M_{X'_1, \psi'_1}, \Delta_{X, \psi}/M_{X'_1, \psi'_1})$ 
in the middle panels, we vary $ \epsilon = 0.001, 0.01, 0.1 $. We can find $ \Delta_{X, \psi}/M_{X'_1, \psi'_1} < 0.5 $ for the whole parameter space which is consistent with the co-annihilation scenario.  On the other hand,  the allowed DM mass range shrinks when 
$ \epsilon $ is decreasing.   
Finally, we fix $ \Delta_{X, \psi} = 0.1 M_{X'_1, \psi'_1}$ in the right panels and show the lines which can satisfy $ \Omega_{X'_1, \psi'_1}h^2 = 0.12 $. In the parameter space $(m_{A'}, \epsilon)$, we vary $ M_{X'_1, \psi'_1} = 0.1, 1, 10 $ GeV. When $ M_{X'_1, \psi'_1} + M_{X'_2, \psi'_2}\sim m_{A'} $,  there is a deep gorge which comes from the resonant annihilation.  In this resonant region,  much smaller $ \epsilon $ values are allowed.

\section{Constraints from accelerators and cosmology}\label{Sec:constraints}

Taking $q_X = -4$ for both scalar and fermion C2CDM models as examples, we already presented the constraints from relic density and DM direction detection in the previous section. In this section, we focus on other constraints from accelerators and cosmology, especially from the collider searches. 

\begin{figure}
\centering
\includegraphics[width=2.8in] {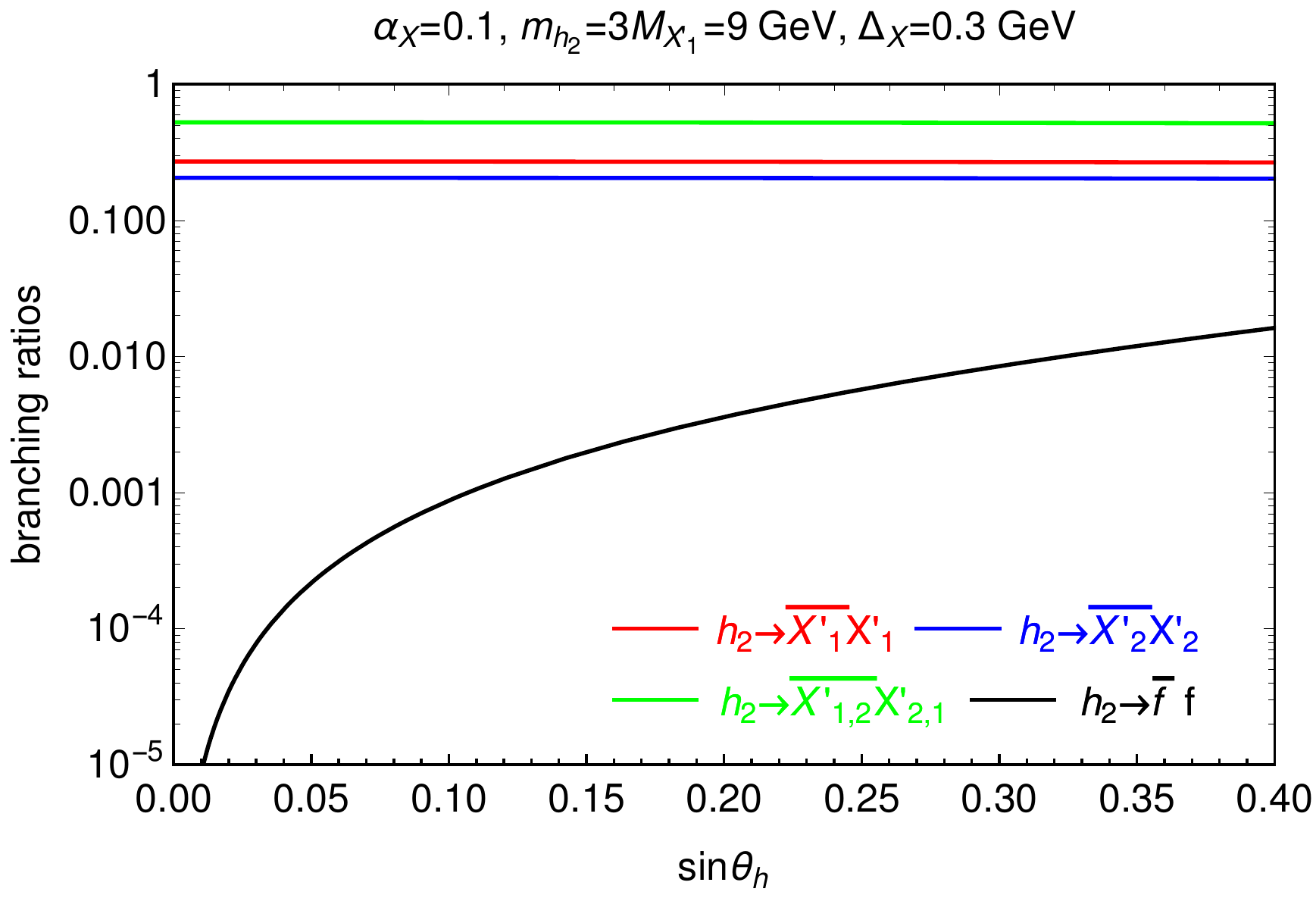} 
\includegraphics[width=2.8in]{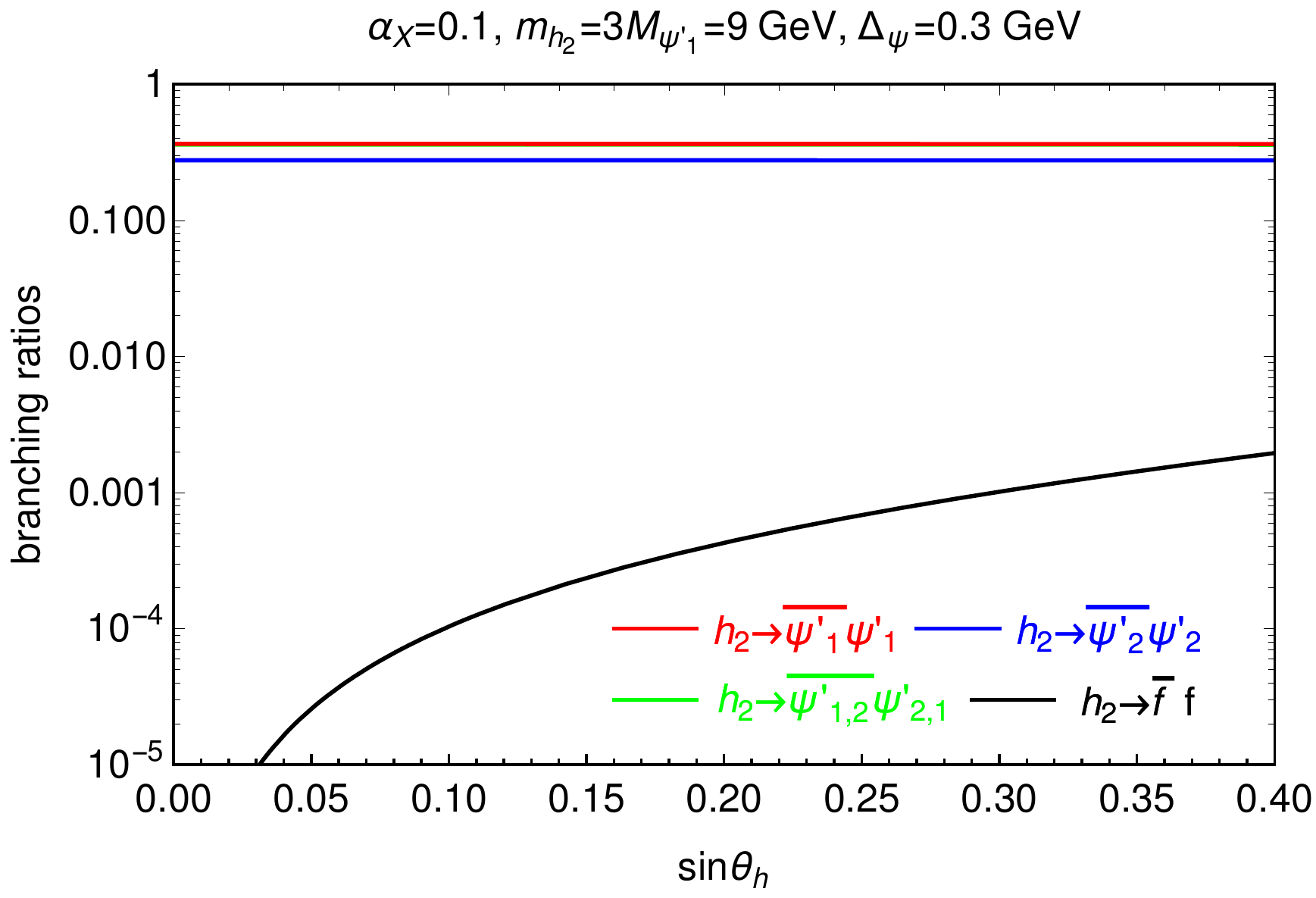} 
\caption{ 
The $h_2$ decay branching ratios for varying $\sin\theta_h$ with $\alpha_X = 0.1$, $m_{h_2} = 3 M_{X'_1,\psi'_1} = 9$ GeV, $\Delta_{X,\psi} = 0.3$ GeV. 
}\label{fig:BR_h2}
\end{figure}

\begin{figure}
\centering 
\includegraphics[width=2.8in]{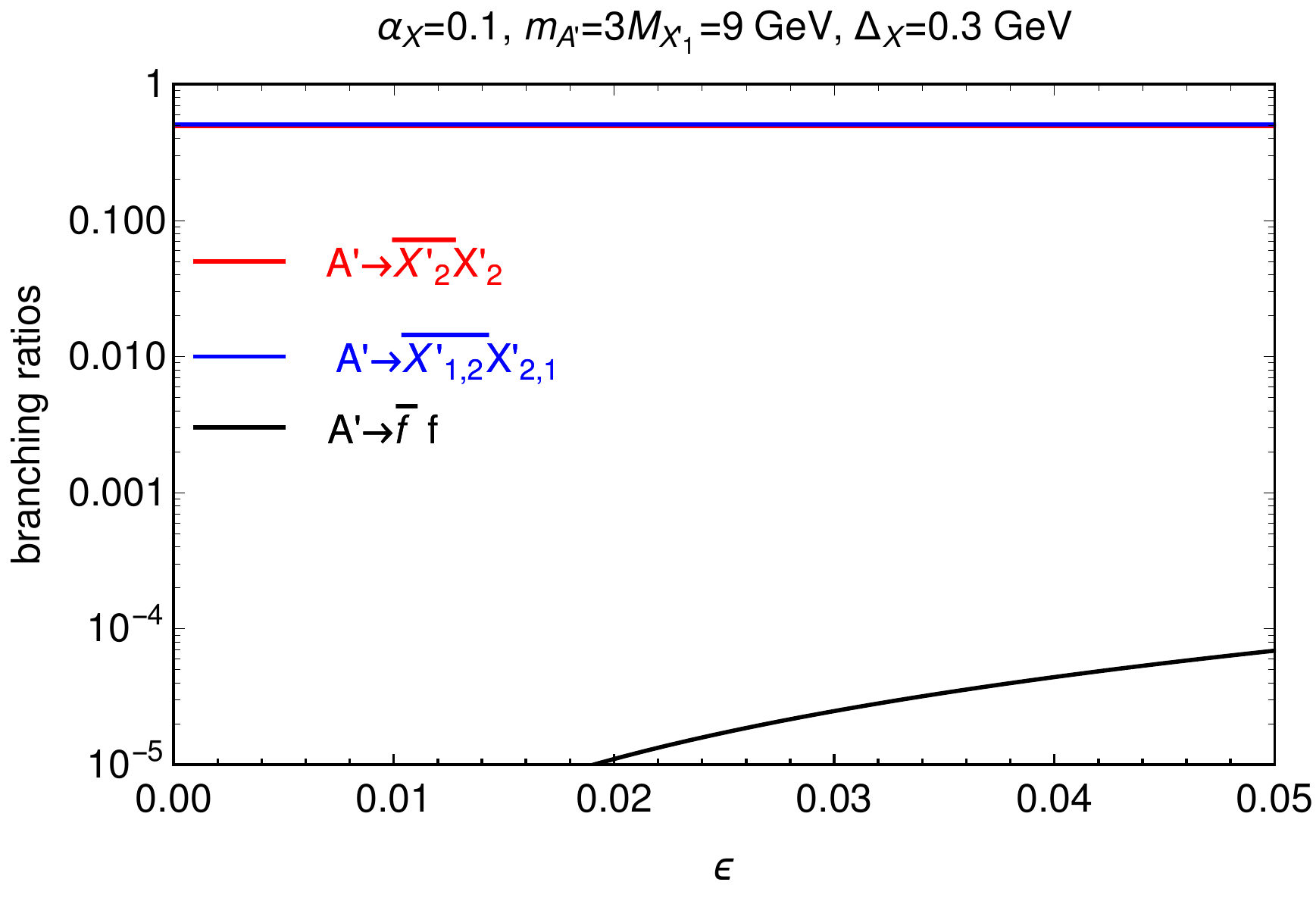} 
\includegraphics[width=2.8in]{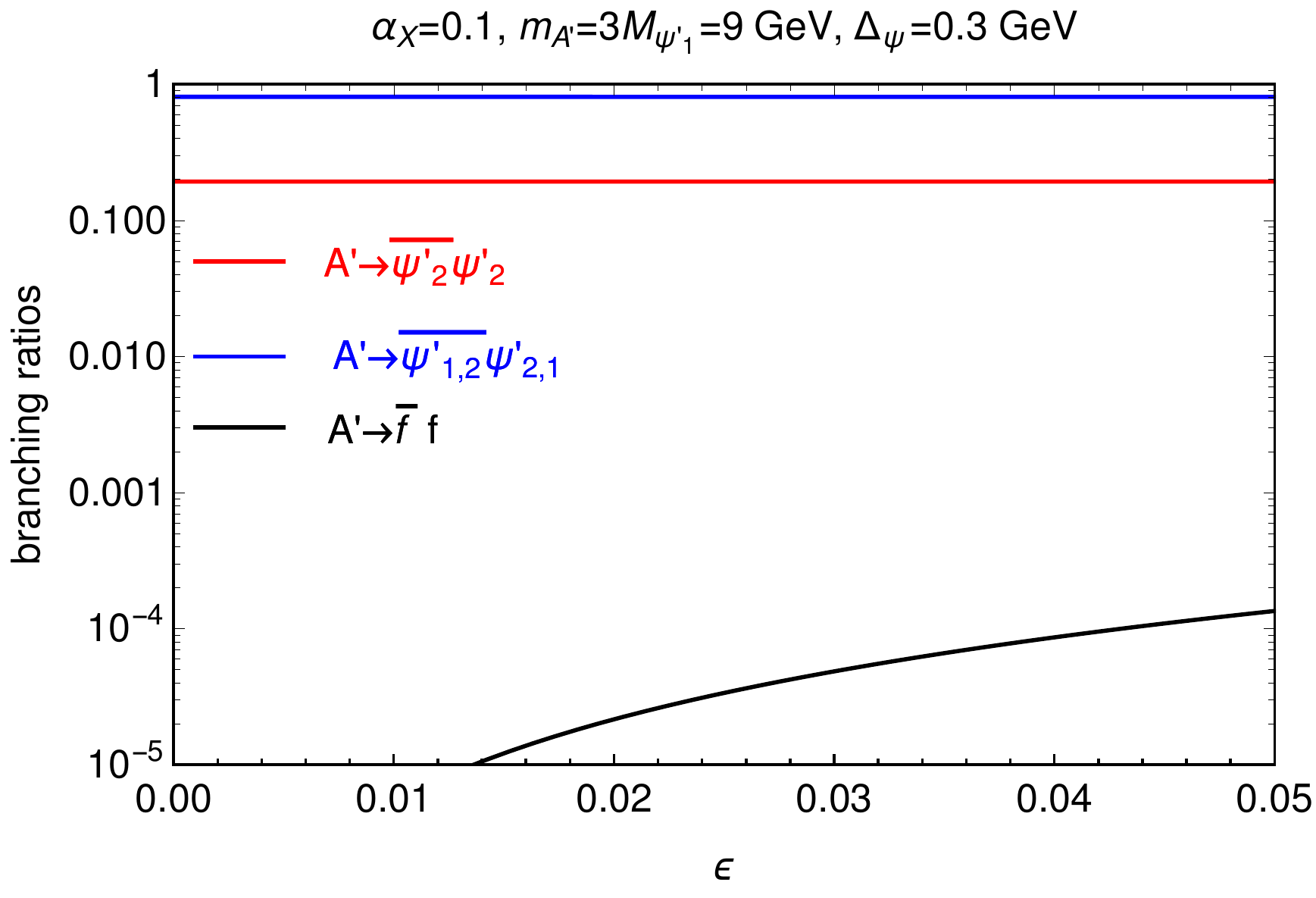}
\caption{ 
The $A'$ decay branching ratios for varying $\epsilon$ with $\alpha_X = 0.1$, $m_{A'} = 3 M_{X'_1,\psi'_1} = 9$ GeV, $\Delta_{X,\psi} = 0.3$ GeV.
}\label{fig:BR_Ap}
\end{figure}

Before moving to each constraint, we first study the decay patterns and branching ratios of $h_2$ and $A'$ which are important for the collider search strategies for these two new particles. 
For the concrete examples, we first fix the model parameters $\alpha_X = 0.1$, $m_{A',h_2} = 3M_{X'_1,\psi'_1}$ GeV, and $\Delta_{X,\psi} = 0.1$ GeV for scalar and fermionic C2CDM models. 
We choose $M_{X'_1,\psi'_1} = 3$ GeV as one example, and show the $h_2$ decay 
branching ratios with $\sin\theta_h$ in the upper panel of Fig.~\ref{fig:BR_h2}, and $A'$ 
decay branching ratios with $\epsilon$ in the lower panel of Fig.~\ref{fig:BR_Ap}, 
respectively. On the other hand, we choose $\sin\theta_h = \epsilon = 10^{-2}$ as the 
other example and show the $h_2$ and $A'$ decay branching ratios with respective to $M_{X'_1,\psi'_1}$ in the upper and lower panels in Fig.~\ref{fig:BR_Ap}, respectively. 
Note we closely follow Ref.~\cite{Liu:2014cma} for $h_2$ visible decays via the mixing between $h$ and $h_X$ and apply Eqs.~(\ref{Eq:XXh}) and~(\ref{Eq:psipsih}) to calculate $h_2\rightarrow X'_1 X'^{\ast}_1 (\psi'_1 \overline{\psi'_1})$ for invisible decays and $h_2\rightarrow X'_1 X'^{\ast}_2 (\psi'_1 \overline{\psi'_2})$,  $h_2\rightarrow X'_2 X'^{\ast}_2 (\psi'_2 \overline{\psi'_2})$ for semi-visible decays. Similarly, we follow Ref.~\cite{Liu:2014cma} for $A'$ visible decays via the kinematic mixing and apply Eq.~(\ref{Eq:XXAp}) and~(\ref{Eq:psipsiAp}) to calculate $A'\rightarrow X'_1 X'^{\ast}_1 (\psi'_1 \overline{\psi'_1})$ for invisible decays and $A'\rightarrow X'_1 X'^{\ast}_2 (\psi'_1 \overline{\psi'_2})$, $A'\rightarrow X'_2 X'^{\ast}_2 (\psi'_2 \overline{\psi'_2})$ for semivisible decays. 
It's clear to see that both $h_2$ and $A'$ mainly decay to dark sector particles in our interested parameter space. 

\begin{figure}
\centering
\includegraphics[width=3.2in]{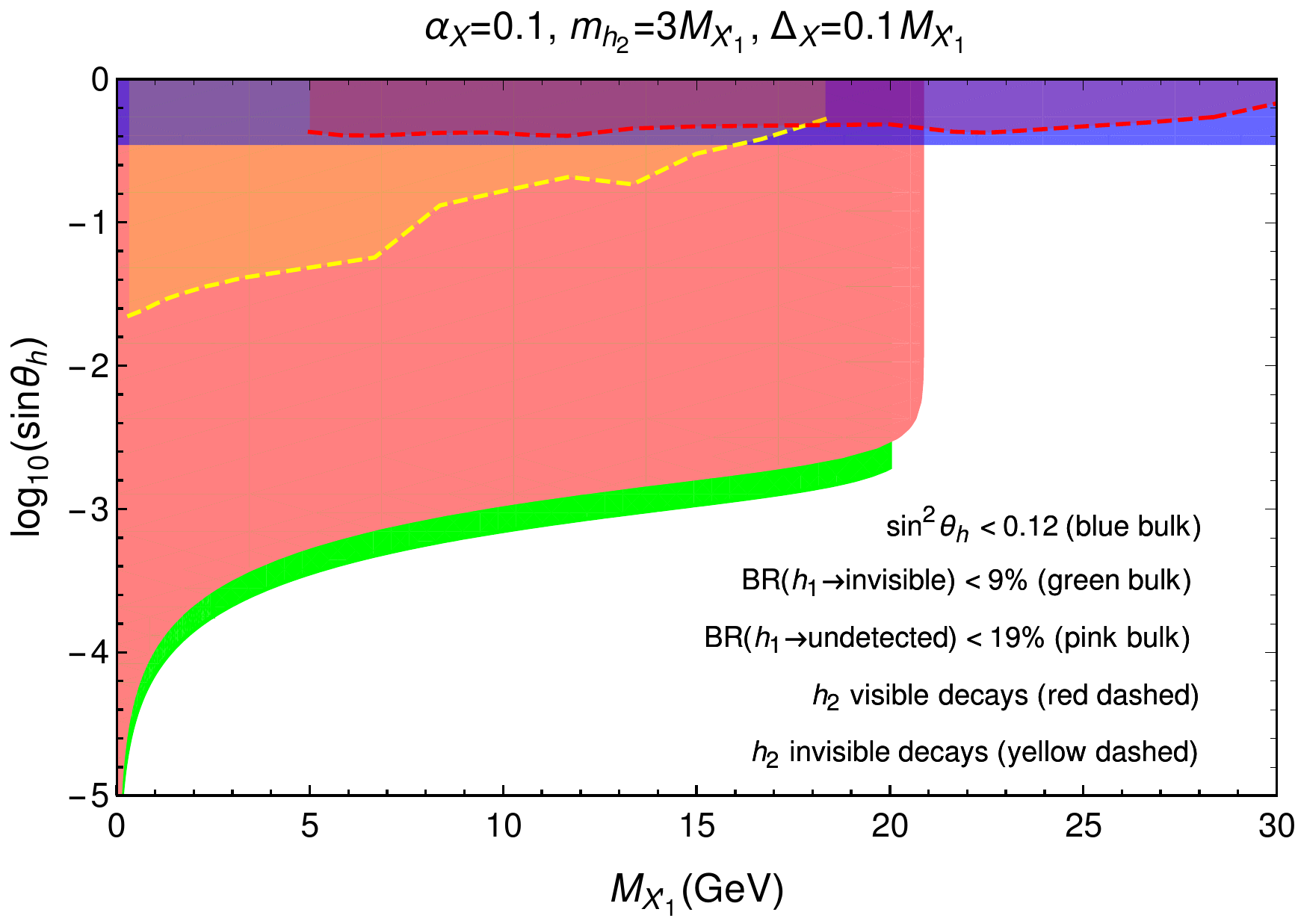} 
\includegraphics[width=3.2in]{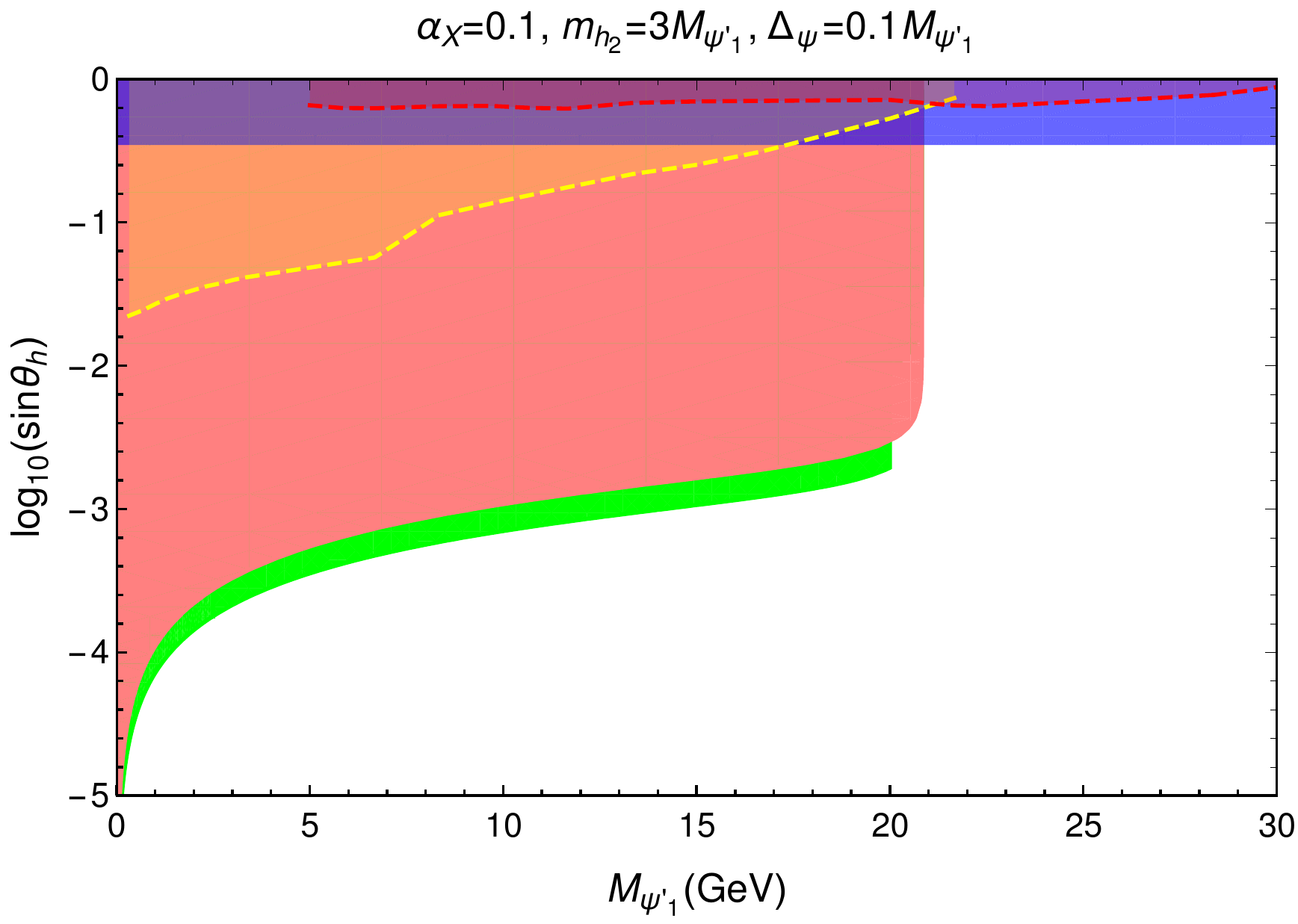}
\caption{ 
The allowed parameter space of ($M_{X'_1}$, $\sin\theta_h$) (left) and ($M_{\psi'_1}$, $\sin\theta_h$) (right) with the following constraints : $\sin^2\theta_h < 0.12$~\cite{Aad:2015pla,Khachatryan:2016vau} (blue bulk), $ BR(h_1\rightarrow\text{invisible}) < 9\% $~Ref.~\cite{ATLAS:2020qdt} (green bulk), $ BR(h_1\rightarrow\text{undetected}) < 19\% $~\cite{ATLAS:2020qdt} (pink bulk), $h_2$ visible decays~\cite{LEPWorkingGroupforHiggsbosonsearches:2003ing} (red dashed line) and $h_2$ invisible decays~\cite{OPAL:2007qwz} (yellow dashed line).
}\label{fig:h2_cons}
\end{figure}

\begin{figure}
\centering
\includegraphics[width=3.2in]{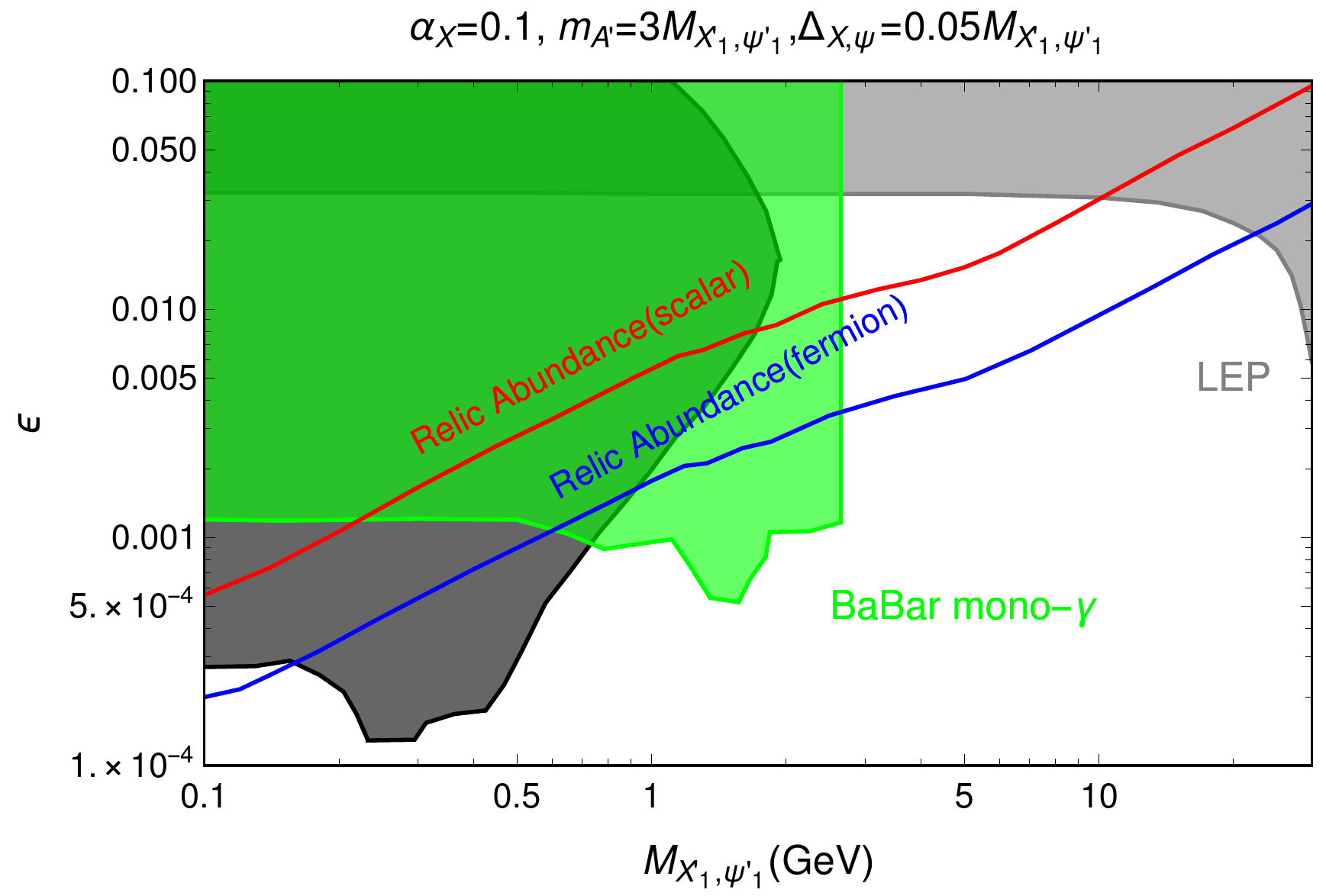} 
\includegraphics[width=3.2in]{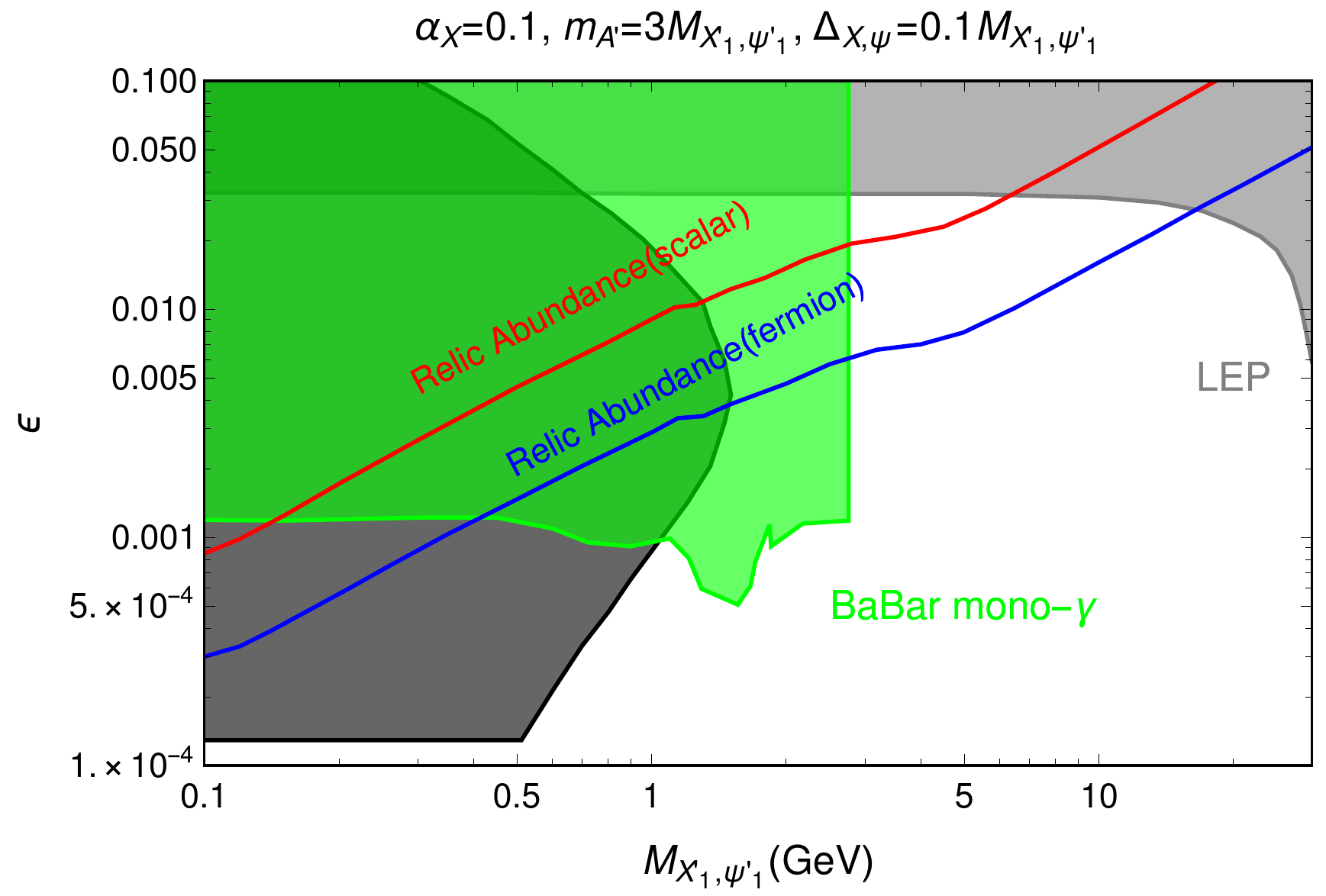}
\caption{ 
The same as the left panel in Fig.~\ref{fig:2DM_con} except for $\Delta_{X,\psi} = 0.05 M_{X'_1,\psi'_1}$ (left) and $\Delta_{X,\psi} = 0.1 M_{X'_1,\psi'_1}$ (right). 
}\label{fig:Ap_cons}
\end{figure}

We then divide related constraints in both scalar and fermionic C2CDM models with the following five categories : 
\begin{itemize} 
\item \textbf{The mixing angle $ \theta_h $ between $h$ and $h_X$} \\ 
According to the LHC Higgs boson measurements, the mixing angle $ \theta_h $ between $h$ and $h_X$ is constrained as $\sin^2\theta_h < 0.12 $ at $ 95\% $ C.L.~\cite{Aad:2015pla,Khachatryan:2016vau}. We show this constraint in Fig.~\ref{fig:h2_cons} with the blue bulk.  
\item \textbf{SM-like Higgs boson $h_1$ invisible and exotic decays} \\ 
In our C2CDM models, we have  $h_1\rightarrow X'_1 X'^{\ast}_1 (\psi'_1 \overline{\psi'_1})$ for the invisible decay and $h_1\rightarrow X'_1 X'^{\ast}_2 (\psi'_1 \overline{\psi'_2})$,  $h_1\rightarrow X'_2 X'^{\ast}_2 (\psi'_2 \overline{\psi'_2})$ for exotic decays. 
On the other hand, if $ m_{h_2, A'} < m_{h_1}/2 $, we also involve $h_1\rightarrow A' A'\rightarrow X'_1 X'^{\ast}_1 X'_1 X'^{\ast}_1 (\psi'_1 \overline{\psi'_1} \psi'_1 \overline{\psi'_1}) $ and $h_1\rightarrow h_2 h_2\rightarrow X'_1 X'^{\ast}_1 X'_1 X'^{\ast}_1 (\psi'_1 \overline{\psi'_1} \psi'_1 \overline{\psi'_1}) $ for invisible decays. Similarly, there are other exotic decay channels for $ m_{h_2, A'} < m_{h_1}/2 $, including $h_1\rightarrow A' A'\rightarrow X'_i X'^{\ast}_j X'_k X'^{\ast}_l (\psi'_i \overline{\psi'_j} \psi'_k \overline{\psi'_l}) $ and $h_1\rightarrow h_2 h_2\rightarrow X'_i X'^{\ast}_j X'_k X'^{\ast}_l (\psi'_i \overline{\psi'_j} \psi'_k \overline{\psi'_l}) $ where $ i, j, k, l = 1,2 $ except for $ i=j=k=l=1 $.
We take $ BR(h_1\rightarrow\text{invisible}) < 9\% $ and $ BR(h_1\rightarrow\text{undetected}) < 19\% $ at $ 95\% $ C.L. as reported in Ref.~\cite{ATLAS:2020qdt} for the green and pink bulks in Fig.~\ref{fig:h2_cons}, respectively. 
Notice that once $ \Delta_{X,\psi}\lesssim 2 $ GeV, the soft visible objects cannot be identified at the LHC~\cite{ATLAS:2019lng} and we classify them to be constrained from $h_1$ invisible decays.    
\item \textbf{$ h_2 $ visible and invisible decays} \\ 
We have $h_2\rightarrow X'_1 X'^{\ast}_1 (\psi'_1 \overline{\psi'_1})$ for invisible decays and $h_2\rightarrow X'_1 X'^{\ast}_2 (\psi'_1 \overline{\psi'_2})$,  $h_2\rightarrow X'_2 X'^{\ast}_2 (\psi'_2 \overline{\psi'_2})$ for semivisible decays. 
As we already see that $ \sin\theta_h $ is severely constrained to be very small from $h_1$ invisible and exotic decays, so we ignore the constraints from BaBar~\cite{BaBar:2010eww,BaBar:2013npw}, Belle~\cite{Belle:2017oht,Belle:2018pzt} and BESIII~\cite{BESIII:2020sdo} experiments for light $ h_2 $ invisible decays which are much weaker than the above ones. We only include constraints from $ h_2 $ visible and invisible decays at LEP~\cite{LEPWorkingGroupforHiggsbosonsearches:2003ing,OPAL:2007qwz} as shown in the red and yellow dashed lines of Fig.~\ref{fig:h2_cons}, separately. Again, for $ \Delta_{X,\psi}\lesssim 2 $ GeV~\cite{OPAL:2007qwz}, we classify them to be constrained from $h_2$ invisible decays.

\item \textbf{$A'$ invisible decays} \\ 
Again, we have $A'\rightarrow X'_1 X'^{\ast}_1 (\psi'_1 \overline{\psi'_1})$ for invisible decays and $A'\rightarrow X'_1 X'^{\ast}_2 (\psi'_1 \overline{\psi'_2})$, $A'\rightarrow X'_2 X'^{\ast}_2 (\psi'_2 \overline{\psi'_2})$ for semivisible decays.  
As we already see that $ \epsilon $ is severely constrained to be very small such that $ A' $ will dominantly decay to dark sector particles and its visible decay constraints are ignorable. 
We remain the novel $ A' $ semivisible decays in the next section and take into account the invisible decay constraints from BaBar~\cite{BaBar:2017tiz} and LEP~\cite{Hook:2010tw} experiments in green and gray bulks of Fig.~\ref{fig:Ap_cons}, separately. 
Finally, we take into account the relevant beam dump experimental bounds from LSND~\cite{deNiverville:2011it}, MiniBoonNE~\cite{MiniBooNE:2017nqe} and also non-observation of $X'_2$, $\psi'_2$ decays in E137~\cite{Berlin:2018pwi}, NuCal and CHARM~\cite{Tsai:2019buq} in the black bulk of Fig.~\ref{fig:Ap_cons}.

\item \textbf{The decay lengths of $X'_2$, $\psi'_2$ and the lower bound of $M_{X'_1, \psi'_1}$} \\ 
Since we require $ \Delta_{X,  \psi} > 2 m_e $ in our numerical analysis for the single-component DM, the lifetimes of $ X'_2 $, $ \psi'_2 $ are much smaller than $\sim {\cal O}(1)$ sec and free from cosmological constraints.    
However, for $\Delta_{X, \psi} < 2 m_e $, it's interesting to study how small the mass splitting $\Delta_{X, \psi}$ can make the C2CDM models become the two-component DM scenario. 
In this case,  $X'_2$ ($\psi'_2$) mainly decays into $X'_1$ ($\psi'_1$) and neutrino pair. For example, we found at $ \Delta_{X, \psi} < 2m_e$, $m_{A^\prime}\gtrsim {\cal O}(100)\, {\rm MeV}$, $\epsilon = 10^{-4}$ and 
$\alpha_X = 0.1$, the lifetimes of $ X'_2 $, $ \psi'_2 $ become much longer than the age 
of universe ($\sim 4.3 \times 10^{17} \, {\rm sec}$). It is because of the destructive interference between two decay channels mediated by $Z$ boson and the dark photon~\cite{Batell:2009vb}. 
The decay rate of $X^\prime_2$  ($\psi^\prime_2$) into $X^\prime_1$ ($\psi^\prime_1$) and neutrino pair mediated by Z boson and the dark photon is given by,
\begin{equation}
\Gamma_{X^\prime_2 \, (\psi^\prime_2) \rightarrow X^\prime_1 \, (\psi^\prime_1) \, \nu \overline{\nu}} \simeq \frac{\alpha_e \alpha_X (1-q_X)^2}{315 \, \pi} \frac{\Delta^9_{X,\psi}}{m^4_{A'} \, m^4_Z} \frac{s^2_{2X,2\psi }}{c^2_W} \,  \epsilon^2 ~,
\end{equation}
for three species of neutrinos as decay products, which is consistent with the Eq. (9) of Ref.~\cite{Batell:2009vb} \footnote{Our result is a factor of 2 larger than  the result in Ref.~\cite{Batell:2009vb},   because in our case DM are 
complex scalar and Dirac fermion, while the WIMPs in Ref.~\cite{Batell:2009vb} are 
real scalar and Majorana fermion.}.
Hence, $ X'_2 $, $ \psi'_2 $ are long-lived enough and can be DM candidates as well.  
On the other hand, the BBN constraint~\cite{Krnjaic:2019dzc} shows the lower bounds of $M_{X'_1}\gtrsim 6$ MeV and $M_{\psi'_1}\gtrsim 9$ MeV, respectively.      
\end{itemize}

Finally, we show the ($M_{X'_1,\psi'_1}$, $\sin\theta_h$) parameter space and relevant constraints in Fig.~\ref{fig:h2_cons} with $m_{A',h_2} = 3M_{X'_1,\psi'_1} $ and $\Delta_{X,\psi} = 0.1 M_{X'_1,\psi'_1}$. 
Since we have fixed the large coupling $g_X$ with $\alpha_X = 0.1$, according to Eq.~(\ref{eq:Zpmass}), $v_X$ can be pinned down if $m_{A'}$ and $g_X$ are fixed such that $h_1 h_2 h_2$ coupling can be increased for a large $v_X$. 
On the other hand, the large coupling $g_X$ also increases the $h_1 A' A'$ coupling. Hence, the Higgs boson invisible and exotic decays are the dominant constraint for the kinematically allowed regions $M_{X'_1,\psi'_1} < 21$ GeV. However, for $M_{X'_1,\psi'_1}\geq 21$ GeV, the main constraint from LEP is much weaker and there are still plenty of parameter space. 
Similarly, we show the ($M_{X'_1,\psi'_1}$, $\epsilon$) parameter space and relevant constraints in Fig.~\ref{fig:Ap_cons} with $m_{A',h_2} = 3M_{X'_1,\psi'_1} $ with $\Delta_{X,\psi} = 0.05 M_{X'_1,\psi'_1}$ (left panel) and $\Delta_{X,\psi} = 0.1 M_{X'_1,\psi'_1}$ (right panel). We also display the observed DM relic abundance for both scalar and fermionic C2CDM models. Most of the parameter space has been ruled out for $M_{X'_1,\psi'_1} < 3$ GeV.  However, the heavier $X'_1,\psi'_1$ can still be explored from the displaced vertex searches at Belle II~\cite{Belle-II:2018jsg} and long-lived particles (LLPs) searches at FASER~\cite{FASER:2019aik}, MATHUSLA~\cite{MATHUSLA:2020uve} and SeaQuest~\cite{Berlin:2018pwi}. 
We will study the novel displaced vertex signatures for C2CDM models in the next section.

\section{Displaced vertex signatures at Belle II}\label{Sec:BelleII}

In this section, we will discuss the novel displaced vertex signatures at Belle II which can be complementary to the mono-photon search. 
Similar to the DM excited state in inelastic DM models, if we assume the mass splitting between $\psi'_2 (X'_2)$ and $\psi'_1 (X'_1)$ is small enough and ignore SM particle masses in the final state, the decay width of $\psi'_2$ can be approximated by 
\begin{equation}
\Gamma_{\psi^\prime_{2} \rightarrow \psi^\prime_{1}f\overline{f}} \simeq \frac{\alpha_e \alpha_X (1-q_X)^2}{15 \pi m^4_{A^\prime}} ~\Delta_{\psi}^5 ~c^2_W s^2_{2\psi} \, \epsilon^2 ~. 
\label{Eq:psi2_width}
\end{equation} 
Similarly, the approximated decay width of $X'_2$ is the same as Eq.~(\ref{Eq:psi2_width}) except for the notation $\psi\rightarrow X$. 
It's obvious that $\psi'_2$, $X'_2$ can easily become the LLPs if $\alpha_X$, $\epsilon$, and $\Delta_{\psi, X}$ are small enough. 
The light LLPs can be searched at fixed target experiments~\cite{Izaguirre:2017bqb} and 
B-factories~\cite{Duerr:2019dmv,Duerr:2020muu,Kang:2021oes}. 

\begin{figure}
\centering
\includegraphics[width=5.0in]{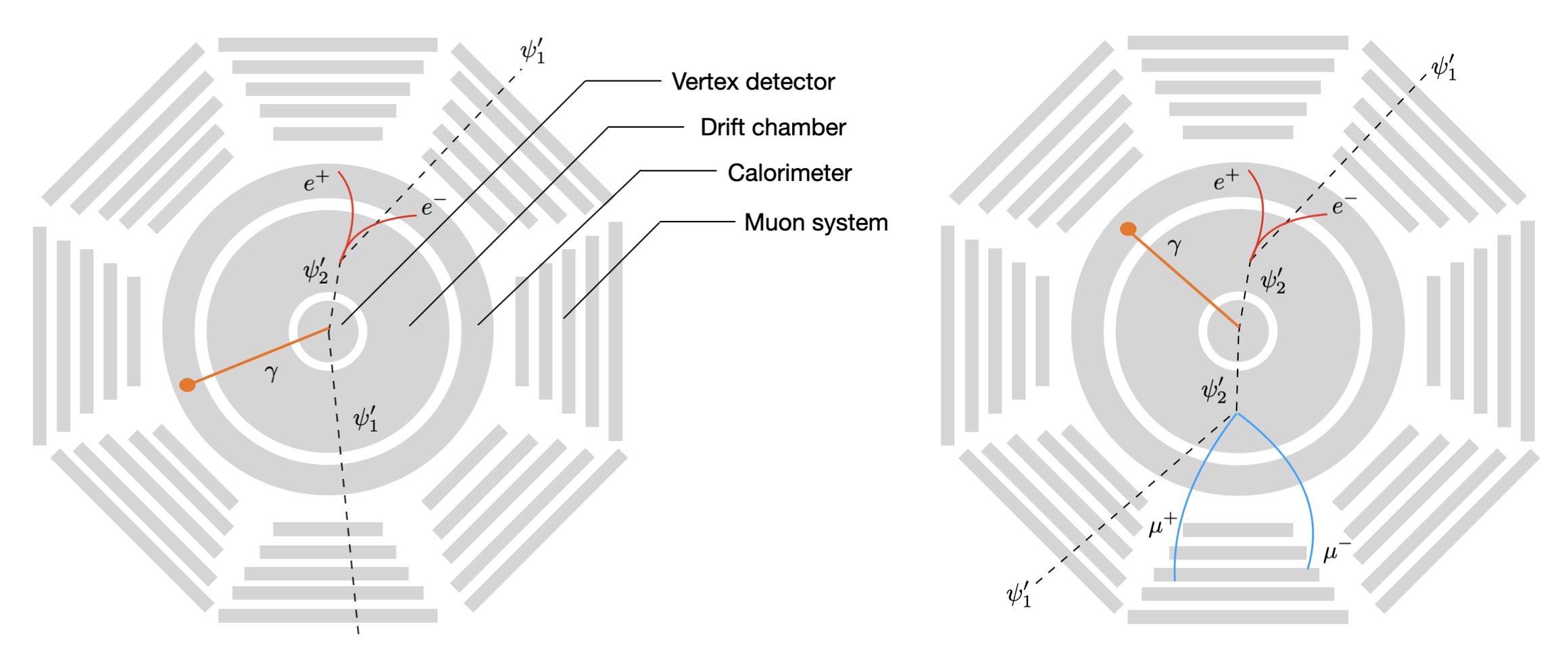} 
\caption{ 
The schematic diagrams for one displaced vertex (left) and two displaced vertices (right) signatures from the fermionic C2CDM models at Belle II.
}\label{fig:signatures}
\end{figure}

As shown in Ref.~\cite{Kang:2021oes}, if we don't explore the DM spin issue of C2CDM models, there is no obvious difference between fermionic and scalar models. Therefore, we will only study the displaced vertex signatures of fermionic C2CDM models in this work. There are four kinds of processes which can be generated at Belle II,  
\begin{equation}
e^+ e^-\rightarrow\psi'_1\overline{\psi'_2}\rightarrow\psi'_1\overline{\psi'_1} (l^+ l^-)       
\end{equation}
\begin{equation}
e^+ e^-\rightarrow\psi'_1\overline{\psi'_2}\gamma\rightarrow\psi'_1\overline{\psi'_1} (l^+ l^-)\gamma       
\end{equation}
\begin{equation}
e^+ e^-\rightarrow\psi'_2\overline{\psi'_2}\rightarrow\psi'_1\overline{\psi'_1} (l^+ l^-) (l^+ l^-)        
\end{equation}
\begin{equation}
e^+ e^-\rightarrow\psi'_2\overline{\psi'_2}\gamma\rightarrow\psi'_1\overline{\psi'_1} (l^+ l^-) (l^+ l^-)\gamma       
\end{equation}
where the dilepton $(l^+ l^-)$ comes from the displaced vertex. Therefore, the signatures can be classified as 
\begin{itemize}
\item (1) One displaced vertex plus missing energy (mET); 
\item (2) The same as (1) but with extra initial state radiation (ISR) photon; 
\item (3) Two displaced vertex plus mET; 
\item (4) The same as (3) but with ISR photon.
\end{itemize} 
We visualize them in Fig.~\ref{fig:signatures}.  
Notice the first two signatures can also be generated from inelastic DM models as shown in Ref.~\cite{Duerr:2019dmv,Kang:2021oes}. However, the signals with two displaced vertices are the unique features in the C2CDM models. 

\begin{figure}
\centering
\includegraphics[width=4.0in]{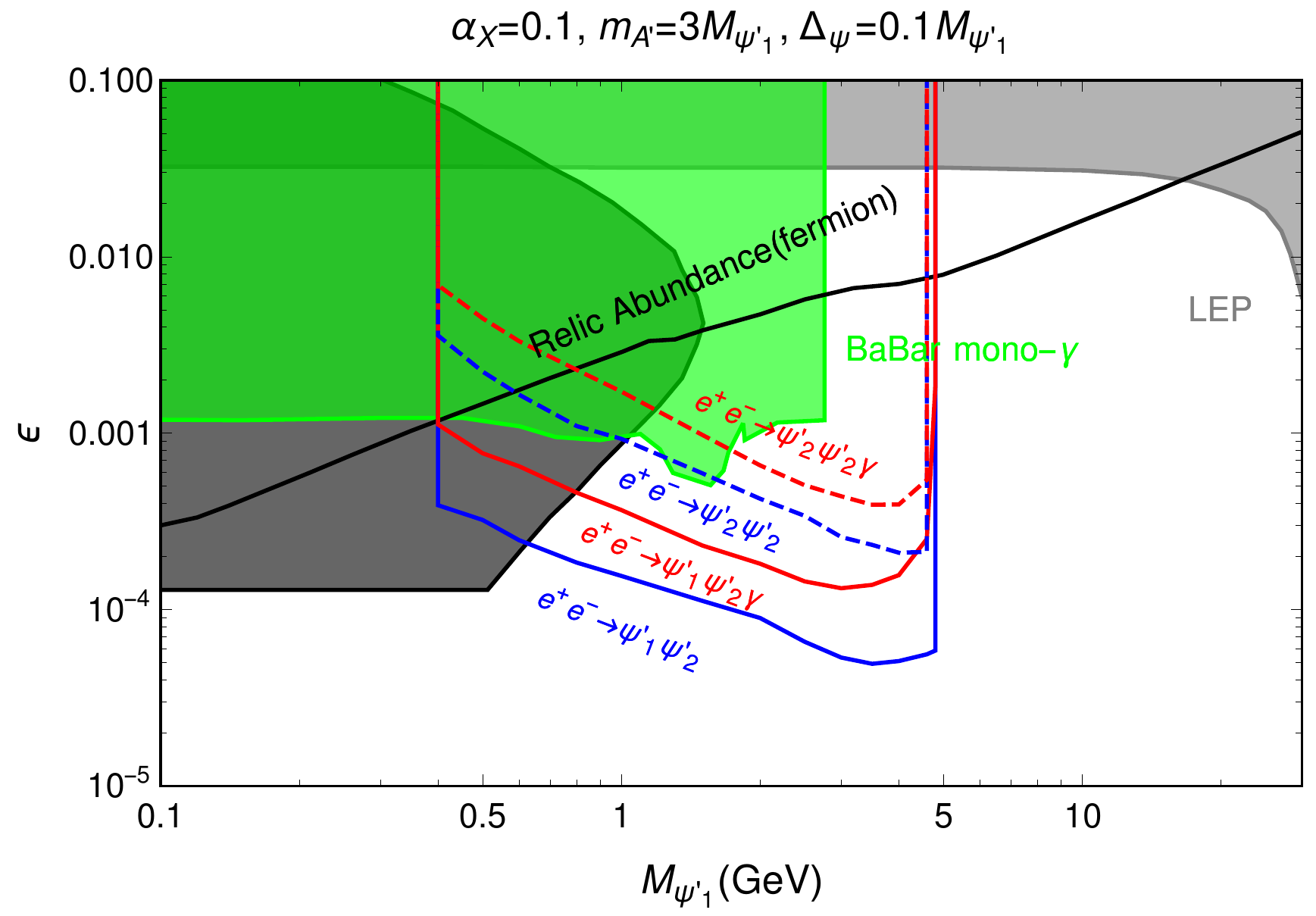} 
\caption{ 
The future bounds from four processes mentioned in the main text with the integrated luminosity of $50$ ab$^{-1}$. Here parameters $\alpha_X = 0.1$, $m_{A'} = 3 M_{\psi'_1}$ and $\Delta_{\psi} = 0.1 M_{\psi'_1}$ are fixed and $ 90\% $ C.L. contours which correspond to an upper limit of $2.3$ events with the assumption of background-free are used. The model-independent LEP bound~\cite{Hook:2010tw}, BaBar mono-$\gamma$ bound~\cite{BaBar:2017tiz} and the observed DM relic abundance line are also shown.
}\label{fig:future_bounds}
\end{figure}

We closely follow Refs.~\cite{Adachi:2018qme,Duerr:2019dmv,Kang:2021oes} for the Belle II detector resolutions and event selections. According to Refs.~\cite{Duerr:2019dmv,Duerr:2020muu,Kang:2021oes}, in regions $0.2 < R_{xy}\leq 0.9 (17.0)$ cm for electron (muon) and $17.0 < R_{xy}\leq 60.0$ cm for both electron and muon with adequate event selections, we can safely assume our signal signatures are background-free. We used the most optimistic integrated luminosity value of $50$ ab$^{-1}$ for the above four processes and displayed future bounds of them in Fig.~\ref{fig:future_bounds}. Here we fix the parameters, $\alpha_X = 0.1$, $m_{A'} = 3 M_{\psi'_1}$ and $\Delta_{\psi} = 0.1 M_{\psi'_1}$, and apply $ 90\% $ C.L. contours which correspond to an upper limit of $2.3$ events with the assumption of background-free. 
On the other hand, we have also added the constraints from model-independent LEP bound~\cite{Hook:2010tw}, BaBar mono-$\gamma$ bound~\cite{BaBar:2017tiz} and the observed DM relic abundance line for the comparison. 
Since the efficiency to detect two displaced vertices at the same time is much smaller than only one displaced vertex, we can expect the bound from the former is weaker than the latter one. However, the signature with two displaced vertices can be used to distinguish our C2CDM models with other BSM models, like inelastic DM models~\cite{Adachi:2018qme,Duerr:2019dmv,Kang:2021oes}, strongly interacting DM models~\cite{Berlin:2018tvf} and Dark seesaw models~\cite{Abdullahi:2020nyr}.

\section{The explanation of $511$ keV $\gamma$-ray line and XENON1T excess}\label{Sec:511X1T}

In the previous sections, we have discussed the C2CDM models including single-component and two-component DM scenarios. For the DM mass larger than $100$ MeV, it can be detected at DM direct detection, indirect detection, fixed target and collider experiments.
In this section, we further extend the DM mass less than ${\cal O}(10)$ MeV which can explain $511$ keV $\gamma$-ray line for the single-component DM scenario, and 
XENON1T excess for the two-component DM scenario.

\subsection{$511$ keV $\gamma$-ray line from the Galactic Center (GC)}
The source of $511$ keV $\gamma$-ray line emission from the Galactic Center (GC) is still 
not confirmed as an evidence of DM. It may come from the undiscovered astrophysical 
objects or light DM annihilation~\cite{Boehm:2003bt}. If the $511$ keV $\gamma$-ray line emission from the GC is explained by the DM origin, the cross 
section for non-self-conjugate DM annihilations into an $e^+ e^-$ pair 
with~\cite{Vincent:2012an} 
\begin{equation}
\langle\sigma v\rangle_{511\gamma}\backsimeq 1.2\times 10^{-31}\left(\frac{M_{\text{DM}}}{\text{MeV}}\right)^2 \frac{\text{cm}^3}{\text{sec}} 
\label{Eq:511Xec}
\end{equation}
are suggested for the NFW + Disk DM density profile. On the other hand, if two $e^+ e^-$ pairs are produced from the DM cascade annihilations to two mediators and then the decay of the mediator into the $e^+ e^-$ pair,  the best fit result for the annihilation cross section in Eq.(\ref{Eq:511Xec}) needs to be divided by a factor of 2. We will study the simple annihilation into $e^+ e^-$ pair for scalar C2CDM models and the cascade annihilation into two $e^+ e^-$ pairs from fermionic C2CDM models in the following. 

\begin{figure}
\centering
\includegraphics[width=3.4in]{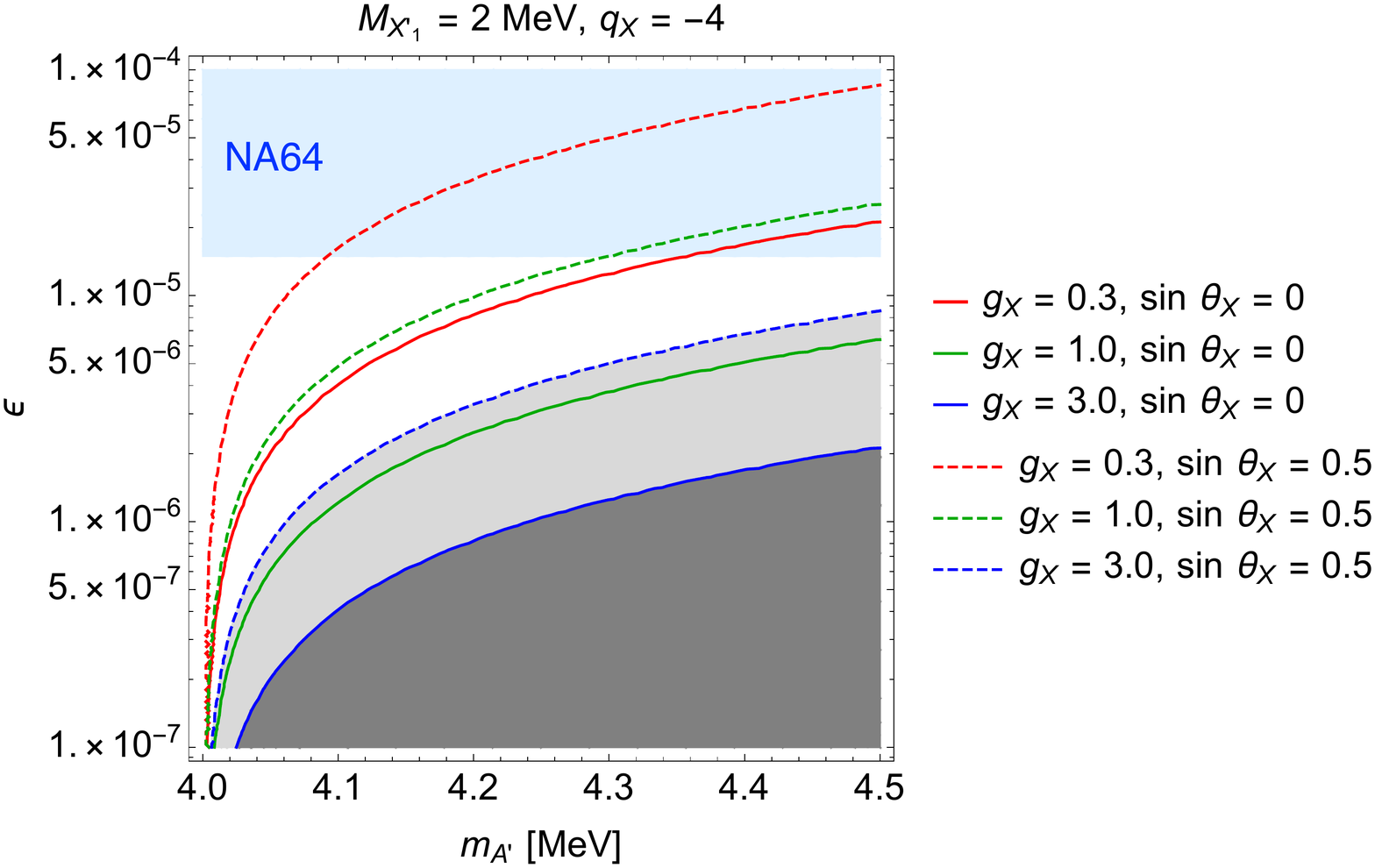} 
\includegraphics[width=2.9in]{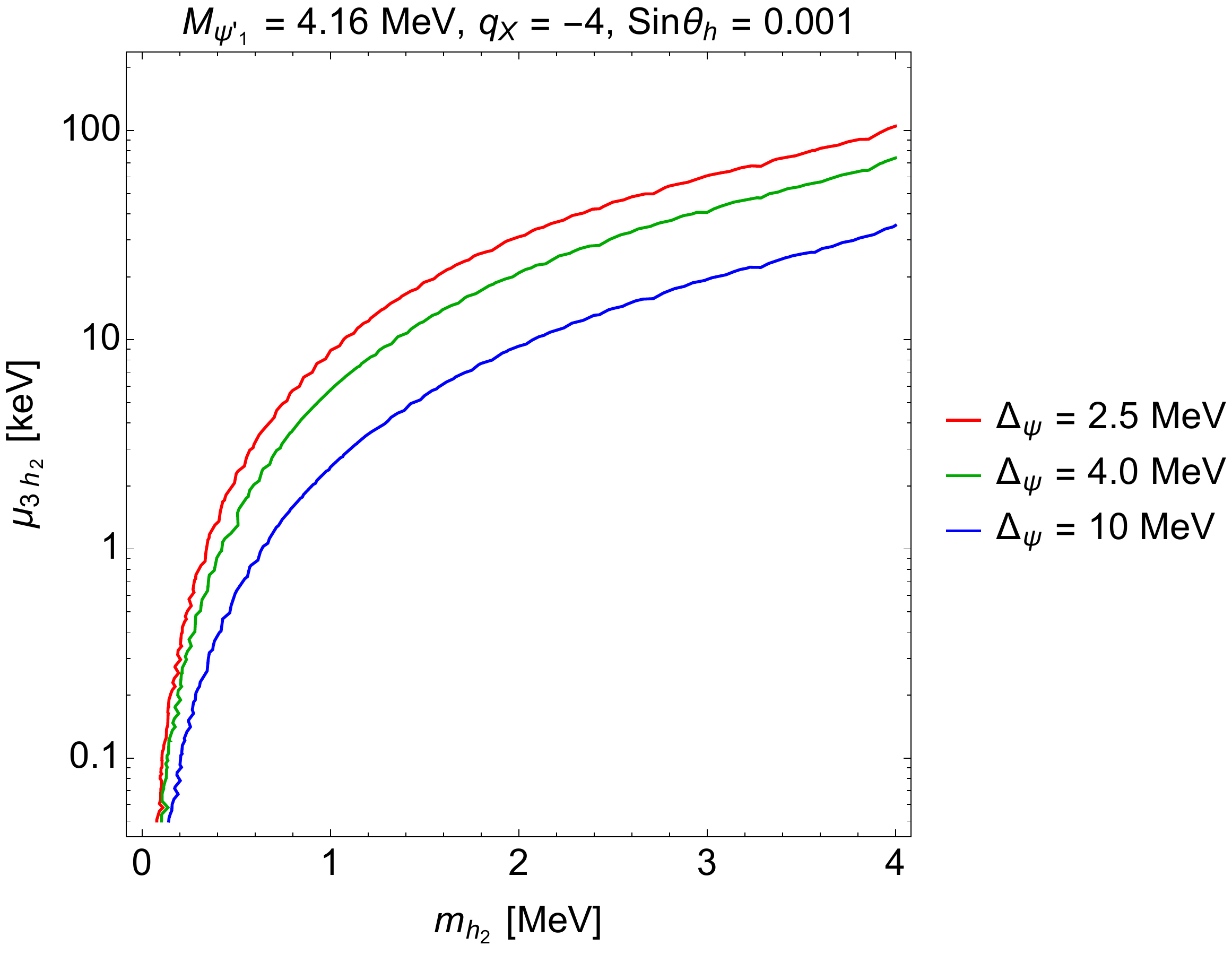}
\caption{ 
The DM interpretation for $511$ keV $\gamma$-ray line in the scalar (left) and fermionic (right) C2CDM models. In the left panel, we fix $M_{X'_1}=2$ MeV, and vary 
$g_X = 0.3$, $1.0$, $3.0$, and $\sin\theta_X = 0.0$, $0.5$ in the $(m_{A'},\epsilon)$ plane. The NA64 constraint~\cite{Banerjee:2019pds} (blue bulk) and the non-perturbative $g_X$ regions (dark and light gray bulks for $\sin \theta_X=0, \, 0.5$ respectively) are also shown.  
In the right panel, we fix $M_{\psi'_1}=4.16$ MeV and 
$\sin\theta_h = 0.001$, and vary $\Delta_{\psi}=2.5$, $4.0$, $10$ MeV in the 
$(m_{h_2},\mu_{3 h_2})$ plane. All lines satisfy the observed DM relic abundance.
}\label{fig:511keV}
\end{figure}

According to Ref.~\cite{Ema:2020fit}, if we try to use DM annihilation with $M_{DM}\lesssim 3$ MeV for $511$ keV $\gamma$-ray line excess issue without conflict of cosmological observations, one proposal is to assume the ratio of neutrino to electron injection $(\sim 1:10^4)$ in the early universe~\cite{Escudero:2018mvt,Sabti:2019mhn}. 
Then the constraints from BBN and CMB on the light DM becomes weaker, because a small fraction of neutrino injection in the early universe can delay neutrino decoupling. This yields the number of effective relativistic degrees of freedom, $N_{\text{eff}}$, 
close to the SM one, $N^{\text{SM}}_{\text{eff}}$.
We will adopt this approach and our C2CDM models can be simply modified by adding extra light dark fermions and a sterile neutrino to enhance the $\nu\overline{\nu}A'$ and $\nu\overline{\nu}h_2$ couplings as the Appendix B in Ref.~\cite{Ema:2020fit} and~\cite{Ko:2014bka}. On the other hand, the stringent CMB constraint for the light DM could be evaded if the dominant annihilation is in the $p$-wave or co-annihilation.

In the scalar C2CDM models, the annihiliation process $X'_1 X'^{\ast}_1\rightarrow e^+ e^-$ is in the p-wave with the following form: 
\begin{equation}
\sigma v(X'_1 X'^{\ast}_1\rightarrow e^+ e^-) = 8\pi\epsilon^2\alpha\alpha^{\prime}_X  c^2_W \frac{M^2_{X'_1} v^2}{3\left[(4M^2_{X'_1}-m^2_{A'})^2 +m^2_{A'}\Gamma^2_{A'}\right]}\left( 1 - \frac{m^2_e}{M^2_{X'_1}} \right)^{3/2}, 
\label{Eq:X1X1ee} 
\end{equation} 
where $\alpha^{\prime}_X = 4\alpha_X (\cos^2\theta_X +q_X\sin^2\theta_X)^2$. If $\Delta_X$ is large enough, the above annihilation process is the dominant one and co-annihilation processes are much suppressed. $X'_2$ is unstable and the lifetime 
$\Gamma_{X'_2\rightarrow X'_1\nu\overline{\nu}}$ is short enough in the cosmological scale\footnote{We have estimated only when $\Delta_X\lesssim 440$ keV, the lifetime of $X'_2$ can be longer than 1 sec.}.   
In addition, we assume $m_{A',h_2} > m_{X'_1}$ such that $X'_1 X'^{\ast}_1\rightarrow A' h_2$  is kinematically forbidden even though it is also $p$-wave process. The Eq.~(\ref{Eq:511Xec}) and Eq.(\ref{Eq:X1X1ee}) can be solved simultaneously to account for both $511$ keV $\gamma$-ray line and the observed DM relic abundance. We fix $M_{X'_1} = 2$ MeV and vary $g_X = 0.3, 1.0, 3.0$ and $\sin\theta_X = 0, 0.5$ as examples to display the allowed parameter space of $(m_{A'},\epsilon)$ plane in the left panel of Fig.~\ref{fig:511keV}. On the other hand, the NA64 constraint~\cite{Banerjee:2019pds} (blue bulk) for the invisible dark photon search and the non-perturbative $g_X$ regions (gray bulk) are also shown.

In the fermionic C2CDM models, the annihilation process $\psi'_1\overline{\psi'_1}\rightarrow e^+ e^-$ is in the s-wave. Therefore, we have to resort to other channels. The first one is the co-annihilation process $\psi'_1 \overline{\psi'_2}\rightarrow e^+ e^-$ in the following form, 
\begin{equation}
\sigma v(\psi'_1\overline{\psi'_2}\rightarrow e^+ e^-) = 4\pi\epsilon^2\alpha\alpha^{\prime\prime}_X c^2_W  \frac{\left[(M_{\psi'_1}+M_{\psi'_2})^2 +2m^2_e\right]}{\left[ m^2_{A'}-(M_{\psi'_1}+M_{\psi'_2})^2\right]^2 +m^2_{A'}\Gamma^2_{A'}}\sqrt{1-\frac{4m^2_e}{(M_{\psi'_1}+M_{\psi'_2})^2}},  
\end{equation}
where $\alpha^{\prime\prime}_X = 4\alpha_X (1-q_X)^2\sin^2\theta_{\psi}\cos^2\theta_{\psi}$.
Note that the assumption of $\sin\theta_{\psi} = 1/\sqrt{1-q_{\psi}}$ is needed to forbid the annihilation process $\psi'_1 \overline{\psi'_1}\rightarrow e^+ e^-$ here. 
The second one is the process
$\psi_1\overline{\psi_1}\rightarrow h_2 h_2$
in the following form, 
\begin{equation}
\begin{split}
\sigma v(\psi'_1\overline{\psi'_1}\rightarrow h_2 h_2) \quad &
\\  =  v^2_{rel} \frac{\sqrt{M^2_{\psi^\prime_1} - m^2_{h_2}}}{384 \, \pi \, M_{\psi^\prime_1}} \Big[ & 16 M^2_{\psi^\prime_1} (M^2_{\psi^\prime_1}-m^2_{h_2})^2 \left( \frac{y^2_{\psi^\prime_1}}{(2 M^2_{\psi^\prime_1}-m^2_{h_2})^2} + \frac{y^2_{\psi^\prime_1 \psi^\prime_2}}{(M^2_{\psi^\prime_1}+M^2_{\psi^\prime_2}-m^2_{h_2})^2} \right)^2
\\ & -8 M_{\psi^\prime_1} (M^2_{\psi^\prime_1}-m^2_{h_2}) \left( \frac{y^2_{\psi^\prime_1}}{(2 M^2_{\psi^\prime_1}-m^2_{h_2})^2} + \frac{y^2_{\psi^\prime_1 \psi^\prime_2}}{(M^2_{\psi^\prime_1}+M^2_{\psi^\prime_2}-m^2_{h_2})^2} \right)
\\ & \quad \quad \times \left( \frac{4 M_{\psi^\prime_1}  y^2_{\psi^\prime_1}}{2 M^2_{\psi^\prime_1}-m^2_{h_2}} + \frac{2 (M_{\psi^\prime_1} +M_{\psi^\prime_2} )y^2_{\psi^\prime_1 \psi^\prime_2}}{M^2_{\psi^\prime_1}+M^2_{\psi^\prime_2} -m^2_{h_2}} + \frac{ y_{\psi^\prime_1} \mu_{3h_2}}{4 M^2_{\psi^\prime_1}-m^2_{h_2}} \right) 
\\ &+ 3 \left( \frac{4 M_{\psi^\prime_1}  y^2_{\psi^\prime_1}}{2 M^2_{\psi^\prime_1}-m^2_{h_2}} + \frac{2 (M_{\psi^\prime_1} +M_{\psi^\prime_2} )y^2_{\psi^\prime_1 \psi^\prime_2}}{M^2_{\psi^\prime_1}+M^2_{\psi^\prime_2}-m^2_{h_2}} + \frac{ y_{\psi^\prime_1} \mu_{3h_2}}{4 M^2_{\psi^\prime_1}-m^2_{h_2}} \right) ^2 \Big] 
\end{split}
\end{equation}
which is p-wave. The notations $y_{\psi^\prime_1}, \ y_{\psi^\prime_1 \psi^\prime_2}$ and $\mu_{3h_2}$ are the coefficients of the operators in the mass basis:
\begin{equation}
\mathcal{L} \supset y_{\psi^\prime_1} h_2 \overline{\psi^\prime_1}\psi^\prime_1 + y_{\psi^\prime_1 \psi^\prime_2 } h_2  ( \overline{\psi^\prime_1} \psi^\prime_2 + h.c. ) -\frac{1}{3!} \mu_{3h_2} h^3_2 ~.
\end{equation} 
And their expressions are given by, 
\begin{equation}
y_{\psi^\prime_1} = \frac{f_\psi}{\sqrt{2}} \sin 2\theta_\psi \cos \theta_h \, , \ y_{ \psi^\prime_1 \psi^\prime_2 } = -\frac{f_\psi}{\sqrt{2}} \cos 2\theta_\psi \cos \theta_h \, , \ \mu_{3h_2} = 3 \left( \frac{c^3_h}{v_X} - \frac{s^3_h}{v} \right) m^2_{h_2} ~.
\end{equation} 
Note that we don't need to assign the specific value for $\sin\theta_{\psi}$ in the second case. The lifetime of $h_2$ is also well below $\sim O(1)$ sec and safe from the BBN constraint.  
Here we focus on the p-wave
 $\psi'_1\overline{\psi'_1}\rightarrow h_2 h_2$
annihilation and assume $\Delta_{\psi}\geq 2.5$ MeV and 
$M_{\psi^\prime_1} = 4.16 {\rm MeV}$  
such that the 
annihilation process can be safely ignored.  
We show the allowed parameter space in the $(m_{h_2},\mu_{3 h_2})$ plane to simultaneously explain both $511$ keV $\gamma$-ray line and the observed DM relic abundance in the right panel of Fig.~\ref{fig:511keV}. 
Here we fix $M_{\psi'_1} = 4.16$ MeV and vary $\Delta_{\psi}= 2.5$, $4.0$, $10$ MeV.

\subsection{XENON1T electron recoil excess}
Finally, the excess of electronic recoil events around 2-3 keV is reported from the XENON1T Collaboration~\cite{XENON:2020rca}. 
This XENON1T excess can also be explained in the two-component DM scenario of C2CDM models.  In this situation, if we ignore the explanation of $511$ keV $\gamma$-ray line, $M_{DM}\geq 10$ MeV is allowed and we don't need to add extra 
light dark fermions and a sterile neutrino to avoid the conflict of cosmological observations. Therefore, if the mass splitting $\Delta_{X,\psi}$ is small enough, the $X'_2$, $\psi'_2$ are also stable in the cosmological scale and can be DM candidates. Furthermore, for $\Delta_{X,\psi}\lesssim 2$ keV, our C2CDM models can explain the XENON1T excess by exothermic scattering of excited DM on  the 
atomic electron in $Xe$ atom, $X'_2 + e_{\rm atomic} \rightarrow X'_1 + e_{\rm free}$, as Ref.~\cite{Harigaya:2020ckz,Lee:2020wmh,Baryakhtar:2020rwy,Bramante:2020zos,Baek:2020owl,Borah:2020smw}. On the other hand, compared with the inelastic DM models, our C2CDM models can have sizable elastic DM-electron scattering cross sections which can be explored in the future DM direct detection experiments. 
However, the dark sector self-interaction $X'_2 X'^{\ast}_2\rightarrow X'_1 X'^{\ast}_1$ ($\psi'_2 \overline{\psi'_2}\rightarrow \psi'_1 \overline{\psi'_1}$) can significantly change the fractions of $X'_2 (\psi'_2)$ and $X'_1 (\psi'_1)$ abundance after the freeze-out as pointed out in~\cite{Harigaya:2020ckz,Lee:2020wmh,Baryakhtar:2020rwy,Bramante:2020zos,Borah:2020smw}. 
Therefore, solving the coupled Boltzmann equations involving this dark sector self-interaction is required, and we leave the complete analysis for the future work.

\section{Discussions and Conclusion}\label{Sec:Conclusion}

The particle nature of dark matter (DM) is still a mystery. As we know, the ordinary matter in the Standard Model (SM) has elaborate structure even its abundance 
in the present Universe is less than $5\%$. The DM abundance is about $5.5$ times larger than the ordinary matter, so we can imagine the structure of dark sector is even richer than the SM sector and DM is not the only particle therein. 
In the SM, the matter field can change its flavor via the charged and neutral current interactions. However, the flavor-changing neutral current (FCNC) interaction is highly suppressed. In this work, we try to ask if the DM field can change its flavor via the FCNC interaction in the dark sector, what kind of models can be built up and what are the unique signatures can be explored ? 

As a prototype of dark flavor-changing neutral current (DFCNC) interaction, the scalar and fermionic crossing two-component dark matter (C2CDM) models with $U(1)_X$ gauge symmetry are built. The function of the dark Higgs field $\Phi$ in C2CDM models is threefold. First, the $U(1)_X$ gauge symmetry is broken via the dark Higgs mechanism of this field and the dark photon becomes massive. Second, this dark Higgs field also communicates between each component of two DM sectors such that the DFCNC interaction between them is induced at tree level. Third, these two DM sectors will mix to each other after $U(1)_X$ gauge symmetry breaking and the mass splitting between them can be generated via the same dark Higgs field.

We then turn to the stability issue of DM candidates in the C2CDM models. We find if $q_X = \frac{3}{2}, \frac{1}{2}, 2, \frac{2}{3}$ in Table~\ref{Tab:U1scoup}, the dim-3 and dim-5 operators that break the global $U(1)$ symmetry can be generated and 
$X'_1$ may not be stable or long-lived enough. Similarly, if $q_X = 2, \frac{1}{2}$ in Table~\ref{Tab:U1fcoup}, the dim-5 operators that break the global $U(1)$ symmetry can be generated and $\psi'_1$ may not be stable. Therefore, we take $q_X = -4$ for both scalar and fermionic C2CDM models as an example which an accidentally residual global $U(1)$ symmetry can make $X'_1$ and $\psi'_1$ stable or long lived enough. 
The allowed parameter space from the relic density and other constraints are studied for the DM mass from $0.1$ to $30$ GeV. 

Because of the DFCNC interactions, there are both diagonal and off-diagonal $X'_1$, $X'_2$ ($\psi'_1$, $\psi'_2$) couplings with dark photon, $A'$, SM-like Higgs boson, $h_1$, and dark Higgs boson, $h_2$ in scalar (fermionic) C2CDM models. 
These interactions can make our C2CDM models be distinguishable from other single-component and two-component DM models. Especially, we study the novel signatures at Belle II with one displaced vertex and two displaced vertices from $e^{+}e^{-}\rightarrow\psi'_1\overline{\psi'_2} (\gamma)$ and $e^{+}e^{-}\rightarrow\psi'_2\overline{\psi'_2} (\gamma)$ processes, respectively. Similarly, there are also multiple displaced vertices from SM-like Higgs boson exotic decays which can be explored at the LHC. 

Finally, we extend our studies to DM mass less than $\sim {\cal O}(10)$ MeV in C2CDM models and try to explain $511$ keV $\gamma$-ray line and XENON1T excess. In the single-component 
DM scenario, we focus on the explanation of $511$ keV $\gamma$-ray line. For the scalar C2CDM models with $M_{X'_1} = 2$ MeV, the p-wave $X'_1 X'^{\ast}_1\rightarrow e^+ e^-$ annihilation process is considered to satisfy both the observed DM relic abundance and $511$ keV $\gamma$-ray line. For the fermionic C2CDM models with $M_{\psi'_1} = 4.16$ MeV, the p-wave $\psi'_1\overline{\psi'_1}\rightarrow h_2 h_2\rightarrow (e^+ e^-)(e^+ e^-)$ cascade annihilation process can also interpret both the observed DM relic abundance and $511$ keV $\gamma$-ray line. Meanwhile, the stringent bound of ${\cal O}(1)$ MeV DM from CMB and BBN can be alleviated by introducing extra 
light dark fermions and a sterile neutrino. 
In the two-component DM scenario, the XENON1T excess can be interpreted with exothermic scattering of excited DM on the atomic electron in $Xe$ atom, $X'_2 (\psi'_2) + e_{\rm atomic} \rightarrow X'_1 (\psi'_1) + e_{\rm free}$ in scalar (fermionic) C2CDM models. Here the small mass splitting, $\Delta_{X,\psi}\lesssim 2$ keV, between $X'_2 (\psi'_2)$ and $X'_1 (\psi'_1)$ is required and the DM mass can be larger than $10$ MeV to evade the cosmological bounds.

\section*{Acknowledgment} 
We would like to thank Joern Kersten and Shu-Yu Ho for useful discussions.
This work is supported by KIAS Individual Grants under Grant No. PG075301 (CTL), and No. PG021403 (PK), and also in part by National Research Foundation of Korea (NRF) Grant No. NRF2019R1A2C3005009 (PK) and No. NRF2021R1A2C1095430 (UM).

\newpage

\appendix

\section{Interactions of dark sectors with the $U(1)_X$ gauge boson and $h_{1,2}$}
\label{Sec:App1}
In this appendix, we collect all the relevant interactions in the dark sector with the $U(1)_X$ gauge boson and 
$h_{1,2}$ for both scalar and fermionic C2CDM models as mentioned in Sec.~\ref{Sec:Model}. 

For the scalar C2CDM models, the $X'_1$, $X'_2$ interaction with $U(1)_X$ gauge boson can be written as
 \begin{equation}
 \begin{split}
  i g_X  C_\mu \Big[ & 2 ( \cos^2 \theta_X + q_X \sin^2 \theta_X )  X'_1 \partial^\mu X'^*_1 +  2 ( \sin^2 \theta_X + q_X \cos^2 \theta_X ) X'_2 \partial^\mu X'^*_2  
\\ &+  (1-q_X) \sin 2\theta_X \left( X'_1 \partial^\mu X'^*_2 + X'_2 \partial^\mu X'^*_1 \right) \Big]
\\ +g^2_X C^2 \Big[ & (\cos^2 \theta_X +q^2_X \sin^2 \theta_X) X'^*_1 X'_1 + (\sin^2 \theta_X +q^2_X \cos^2 \theta_X) X'^*_2 X'_2 
\\ & + \frac{1}{2}(1-q^2_X) \sin 2\theta_X (X'^*_1 X'_2+X'^*_2 X'_1) \Big] ,
 \end{split} 
\label{Eq:XXAp}
 \end{equation}  
and the $X'_1$, $X'_2$ interaction with $h_{1,2}$ are 
\begin{equation}
\begin{split}
& - \left[ \left( \lambda_{HX_1} v c_h + \lambda_{\Phi X_1} v_X s_h \right) c^2_X + \left( \lambda_{HX_2} v c_h + \lambda_{\Phi X_2} v_X s_h \right) s^2_X -\frac{\mu_{X_1X_2\Phi}}{\sqrt{2}} s_h s_{2X} \right] X'^*_1 X'_1 h_1
\\ & - \left[ \left( - \lambda_{HX_1} v s_h + \lambda_{\Phi X_1} v_X c_h \right) c^2_X + \left(- \lambda_{HX_2} v s_h + \lambda_{\Phi X_2} v_X c_h \right) s^2_X -\frac{\mu_{X_1X_2\Phi}}{\sqrt{2}} c_h s_{2X} \right] X'^*_1 X'_1 h_2
\\ & - \left[ \left( \lambda_{HX_1} v c_h + \lambda_{\Phi X_1} v_X s_h \right) s^2_X + \left( \lambda_{HX_2} v c_h + \lambda_{\Phi X_2} v_X s_h \right) c^2_X +\frac{\mu_{X_1X_2\Phi}}{\sqrt{2}} s_h s_{2X} \right] X'^*_2 X'_2 h_1
\\ & - \left[ \left( - \lambda_{HX_1} v s_h + \lambda_{\Phi X_1} v_X c_h \right) s^2_X + \left(- \lambda_{HX_2} v s_h + \lambda_{\Phi X_2} v_X c_h \right) c^2_X +\frac{\mu_{X_1X_2\Phi}}{\sqrt{2}} c_h s_{2X} \right] X'^*_2 X'_2 h_2
\\ & - \left[\frac{1}{2} \left( \lambda_{HX_1} v c_h + \lambda_{\Phi X_1} v_X s_h - \lambda_{HX_2} v c_h - \lambda_{\Phi X_2} v_X s_h \right) s_{2X} + \frac{\mu_{X_1X_2\Phi}}{\sqrt{2}} s_h c_{2X} \right] \left( X'^*_1 X'_2 +X'^*_2 X'_1 \right) h_1
\\ & - \left[\frac{1}{2} \left( - \lambda_{HX_1} v s_h + \lambda_{\Phi X_1} v_X c_h + \lambda_{HX_2} v s_h - \lambda_{\Phi X_2} v_X c_h \right) s_{2X} +\frac{\mu_{X_1X_2\Phi}}{\sqrt{2}} c_h c_{2X} \right] \left( X'^*_1 X'_2 +X'^*_2 X'_1 \right) h_2 ~,
\end{split}
\end{equation}
 
 \begin{equation}
\begin{split}
& - \frac{1}{2} \left[ \left( \lambda_{HX_1} c^2_h + \lambda_{\Phi X_1} s^2_h \right) c^2_X + \left( \lambda_{HX_2} c^2_h + \lambda_{\Phi X_2} s^2_h \right) s^2_X \right] X'^*_1 X'_1 h^2_1
\\ & - \frac{1}{2} \left[ \left( \lambda_{HX_1} s^2_h + \lambda_{\Phi X_1} c^2_h \right) c^2_X + \left( \lambda_{HX_2} s^2_h + \lambda_{\Phi X_2} c^2_h \right) s^2_X \right] X'^*_1 X'_1 h^2_2
\\ & - \frac{1}{2} \left[ \left( - \lambda_{HX_1}+ \lambda_{\Phi X_1} \right) c^2_X + \left( - \lambda_{HX_2} + \lambda_{\Phi X_2} \right) s^2_X \right] s_{2h} X'^*_1 X'_1 h_1 h_2
\\ & - \frac{1}{2} \left[ \left( \lambda_{HX_1} c^2_h + \lambda_{\Phi X_1} s^2_h \right) s^2_X + \left( \lambda_{HX_2} c^2_h + \lambda_{\Phi X_2} s^2_h \right) c^2_X \right] X'^*_2 X'_2 h^2_1
\\ & - \frac{1}{2} \left[ \left( \lambda_{HX_1} s^2_h + \lambda_{\Phi X_1} c^2_h \right) s^2_X + \left( \lambda_{HX_2} s^2_h + \lambda_{\Phi X_2} c^2_h \right) c^2_X \right] X'^*_2 X'_2 h^2_2
\\ & - \frac{1}{2} \left[ \left( - \lambda_{HX_1}+ \lambda_{\Phi X_1} \right) s^2_X + \left( - \lambda_{HX_2} + \lambda_{\Phi X_2} \right) c^2_X \right] s_{2h} X'^*_2 X'_2 h_1 h_2
\\ & - \frac{1}{4} \left[ \left( \lambda_{HX_1} - \lambda_{HX_2} \right) c^2_h + \left( \lambda_{\Phi X_1} - \lambda_{\Phi X_2} \right) s^2_h \right] s_{2X} \left( X'^*_1 X'_2 + X'^*_2 X'_1 \right) h^2_1
\\ & - \frac{1}{4} \left[ \left( \lambda_{HX_1} - \lambda_{HX_2} \right) s^2_h + \left( \lambda_{\Phi X_1} - \lambda_{\Phi X_2} \right) c^2_h \right] s_{2X} \left( X'^*_1 X'_2 + X'^*_2 X'_1 \right) h^2_2
\\ & - \frac{1}{4} \left( - \lambda_{HX_1} + \lambda_{HX_2} + \lambda_{\Phi X_1} - \lambda_{\Phi X_2} \right) s_{2h} s_{2X} \left( X'^*_1 X'_2 + X'^*_2 X'_1 \right) h_1 h_2 ~,
\end{split} 
\label{Eq:XXh}
\end{equation}
 where $s_{h,X} \equiv \sin \theta_{h,X} $, $c_{h,X} \equiv \cos \theta_{h,X} $, $s_{2h,2X} \equiv \sin 2\theta_{h,X} $ and $c_{2h,2X} \equiv \cos 2 \theta_{h,X} $.

For the fermionic C2CDM models, the $\psi^\prime_1$, $\psi^\prime_2$ interaction with $U(1)_X$ gauge boson can be written as 
\begin{equation}
\begin{split}
& -g_X \left( \cos^2 \theta_\psi +q_X \sin^2 \theta_\psi \right)\overline{\psi^\prime_1} C\!\!\!\!/ \, \psi^\prime_1
\\ &- g_X\left( \sin^2 \theta_\psi +q_X \cos^2 \theta_\psi \right)\overline{\psi^\prime_2} C\!\!\!\!/ \, \psi^\prime_2
\\ & -g_X \left( 1-q_X \right) \sin \theta_\psi \cos \theta_\psi \left( \overline{\psi^{\prime}_1} C\!\!\!\!/ \, \psi^\prime_2  +\overline{\psi^{\prime}_2} C\!\!\!\!/ \, \psi^{\prime}_1 \right)
\end{split} 
\label{Eq:psipsiAp}
\end{equation}
and the $\psi^\prime_1$, $\psi^\prime_2$ interaction with $h_{1,2}$ are 
\begin{equation}
\begin{split}
& \frac{f}{\sqrt{2}} \sin 2\theta_\psi  \left( h_1 \sin \theta_h + h_2 \cos \theta_h \right)\overline{\psi^\prime_1} \psi^\prime_1 
\\ & -\frac{f}{\sqrt{2}} \sin 2\theta_\psi \left( h_1 \sin \theta_h + h_2 \cos \theta_h \right) \overline{\psi^\prime_2} \psi^\prime_2
\\ & -\frac{f}{\sqrt{2}} \cos 2\theta_\psi \left( h_1 \sin \theta_h + h_2 \cos \theta_h \right) \left( \overline{\psi^\prime_1} \psi^\prime_2 +\overline{\psi^\prime_2} \psi^\prime_1  \right)
\end{split} 
\label{Eq:psipsih}
\end{equation}

\section{Possible decay channels and partial decay widths for $X^{\prime}_1$ and $\psi^{\prime}_1$ with specific $q_X$}\label{Sec:App2} 

\begin{figure}
\centering
\includegraphics[width=1.1in]{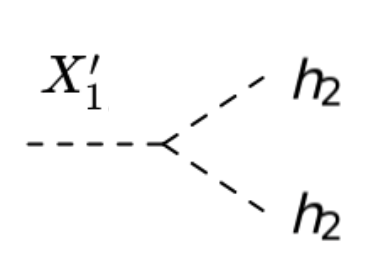}
\includegraphics[width=1.5in]{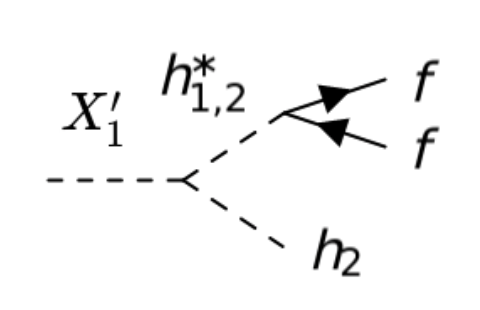}
\includegraphics[width=1.6in]{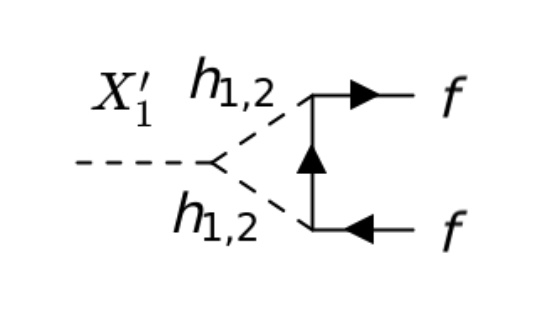}
\includegraphics[width=1.5in]{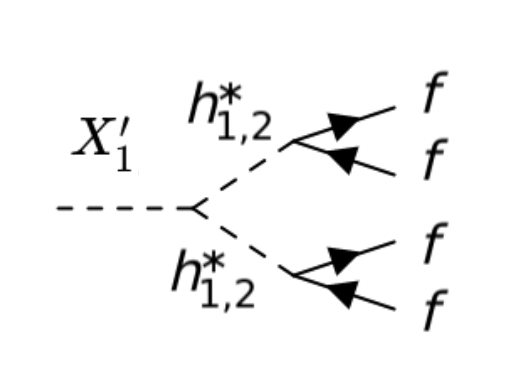}
\caption{ 
Possible $ X^{\prime}_1 $ decay channels in scalar C2CDM models with $ q_X = \frac{3}{2}, \frac{1}{2}, 2, \frac{2}{3} $.
}\label{fig:SC2CDM}
\end{figure}

According to Eq.(\ref{Sdanger3}) and~(\ref{Sdanger5}), we found there are some dangerous dim-3 and dim-5 operastors which can cause the decay of DM candidate $X^{\prime}_1$. Depending on the mass relation between $ X^{\prime}_1 $ and $ h_2 $, we expect possible $ X^{\prime}_1 $ decay channels are 
\begin{enumerate}
\item $ M_{X^{\prime}_1} > 2m_{h_2} $ : $ X^{\prime}_1\rightarrow h_2 h_2 $ 
\item $ m_{h_2}\leq M_{X^{\prime}_1}\leq 2m_{h_2} $ : $ X^{\prime}_1\rightarrow h_2 f\overline{f} $ where $ f $ is the SM fermion. 
\item $ M_{X^{\prime}_1} < m_{h_2} $ : $ X^{\prime}_1\rightarrow f\overline{f} $ (1-loop) and $ X^{\prime}_1\rightarrow f\overline{f} f\overline{f} $. 
\end{enumerate}
where the relevant Feynman diagrams are shown in Fig.~\ref{fig:SC2CDM}. 
The partial decay widths of $ X^{\prime}_1 $ from the above processes are 
\begin{enumerate} 
\item For $M_{X^\prime_1} > 2m_{h_2}$ : 
\begin{equation} 
\Gamma_{X^\prime_1 \rightarrow h_2 h_2 }= \frac{\mu^2}{16\pi M_{X^\prime_1}} \sqrt{ 1- \frac{4m^2_{h_2}}{M^2_{X^\prime_1}} }
\end{equation} 
\item For $m_{h_2}\leq M_{X^{\prime}_1}\leq 2m_{h_2}$ : 
\begin{equation} 
\begin{split}
& \Gamma_{ X^\prime_1 \rightarrow h_2 f \bar{f} } \left( m_f \rightarrow 0 \right) = 
\\ & \quad \frac{\mu^2 y^2}{12 \cdot (4\pi)^3} \frac{M^3_{X^\prime_1}}{m^4_{h_{1,2}}} \left[ \left( 1- \frac{m^2_{h_{1,2}}}{M^2_{X^\prime_1}}\right) \left( 1+  \frac{10m^2_{h_{1,2}}}{M^2_{X^\prime_1}}+\frac{m^4_{h_{1,2}}}{M^4_{X^\prime_1}}\right) -12 \frac{m^2_{h_{1,2}}}{M^2_{X^\prime_1}} \left( 1+\frac{m^2_{h_{1,2}}}{M^2_{X^\prime_1}} \right) \ln \frac{M_{X^\prime_1}}{m_{h_{1,2}}} \right] 
\end{split}
\end{equation}
\item For $ M_{X^\prime_1} < m_{h_2} $ : 
\begin{equation}
\begin{split} 
& \quad \Gamma_{X^\prime_1 \rightarrow f \bar{f} } \left( m_{h_{1,2}} \rightarrow \infty \right) = \frac{9\mu^2 y^4}{2\pi^2 (16\pi)^3} \frac{m^2_f M_{X^\prime_1}}{m^4_{h_{1,2}}} \left( 1-\frac{4m^2_f}{M^2_{X^\prime_1}} \right)^\frac{3}{2},
\\ &  \quad \Gamma_{X^\prime_1 \rightarrow f\bar{f}f\bar{f}} \left( m_{h_{1,2}} \rightarrow \infty,\ m_f \rightarrow 0 \right)= \frac{\mu^2 y^4}{23040 \cdot (2\pi)^5} \frac{M^7_{X^\prime_1}}{m^8_{h_{1,2}}} 
\end{split}
\end{equation} 
\end{enumerate}
where $\mu$ is a general dimension-1 coefficient of the operators $X^\prime_1 h_{1,2} h_{1,2}$, and $y$ is a general Yukawa coupling of the operators $h_{1,2} \bar{f} f $. In the symmetry broken phase, the coefficients of dangerous dim-3 and dim-5 operators are both encoded in $\mu$. 
For simplicity, the interference effects between distinct Feynman diagrams are ignored. 

\begin{figure}
\centering
\includegraphics[width=1.1in]{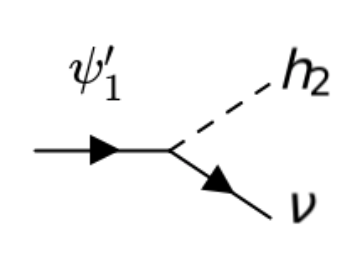}
\includegraphics[width=1.5in]{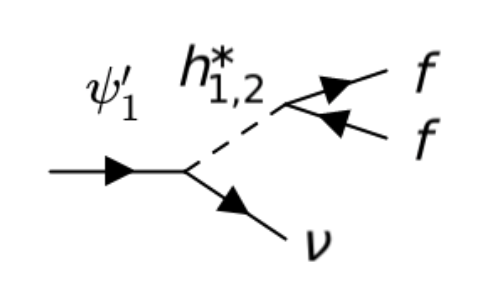}
\caption{ 
Possible $ \psi^{\prime}_1 $ decay channels in fermionic C2CDM models with $ q_X = 2, \frac{1}{2} $.
}\label{fig:FC2CDM}
\end{figure}

Similarly, based on Eq.(\ref{Fdanger5}), we found there are two dangerous dim-5 operators which can cause the decay of DM candidate $\psi^{\prime}_1$. Depending on the mass relation between $ \psi^{\prime}_1 $ and $ h_2 $, we expect the following possible $ \psi^{\prime}_1 $ decay channels 
\begin{enumerate}
\item $ M_{\psi^{\prime}_1} > m_{h_2} $ : $ \psi^{\prime}_1\rightarrow h_2 \nu $ 
\item $ M_{\psi^{\prime}_1}\leq m_{h_2} $ : $ \psi^{\prime}_1\rightarrow f\overline{f}\nu $ where $ f $ is the SM fermion. 
\end{enumerate}
where the relevant Feynman diagrams are shown in Fig.~\ref{fig:FC2CDM}. 
The partial decay widths of $ \psi^{\prime}_1 $ from these two processes are 
\begin{enumerate}
\item For $ M_{\psi^{\prime}_1} > m_{h_2} $ : 
\begin{equation} 
\Gamma_{\psi^{\prime}_1\rightarrow h_2 \nu } \left( m_\nu \rightarrow 0 \right) = \frac{c^2 v^2_X}{64\pi \Lambda^2} M_{\psi^\prime_1} \left( 1- \frac{m^2_{h_{1,2}}}{M^2_{\psi^\prime_1}} \right)^2
\end{equation}
\item For $ M_{\psi^{\prime}_1}\leq m_{h_2} $ : 
\begin{equation} 
\Gamma_{ \psi^{\prime}_1\rightarrow f\overline{f}\nu  } \left( m_\nu, m_f \rightarrow 0 \right) = \frac{c^2 v^2_X y^2_f}{ 192 \cdot (4\pi)^3 \Lambda^2} \frac{ M^5_{\psi^\prime_1}}{m^4_{h_{1,2}}}
\end{equation}
\end{enumerate}

\section{Scalar C2CDM models with non-vanishing $\lambda$'s}\label{Sec:App3} 
\begin{figure}
\centering
\includegraphics[width=3.0in]{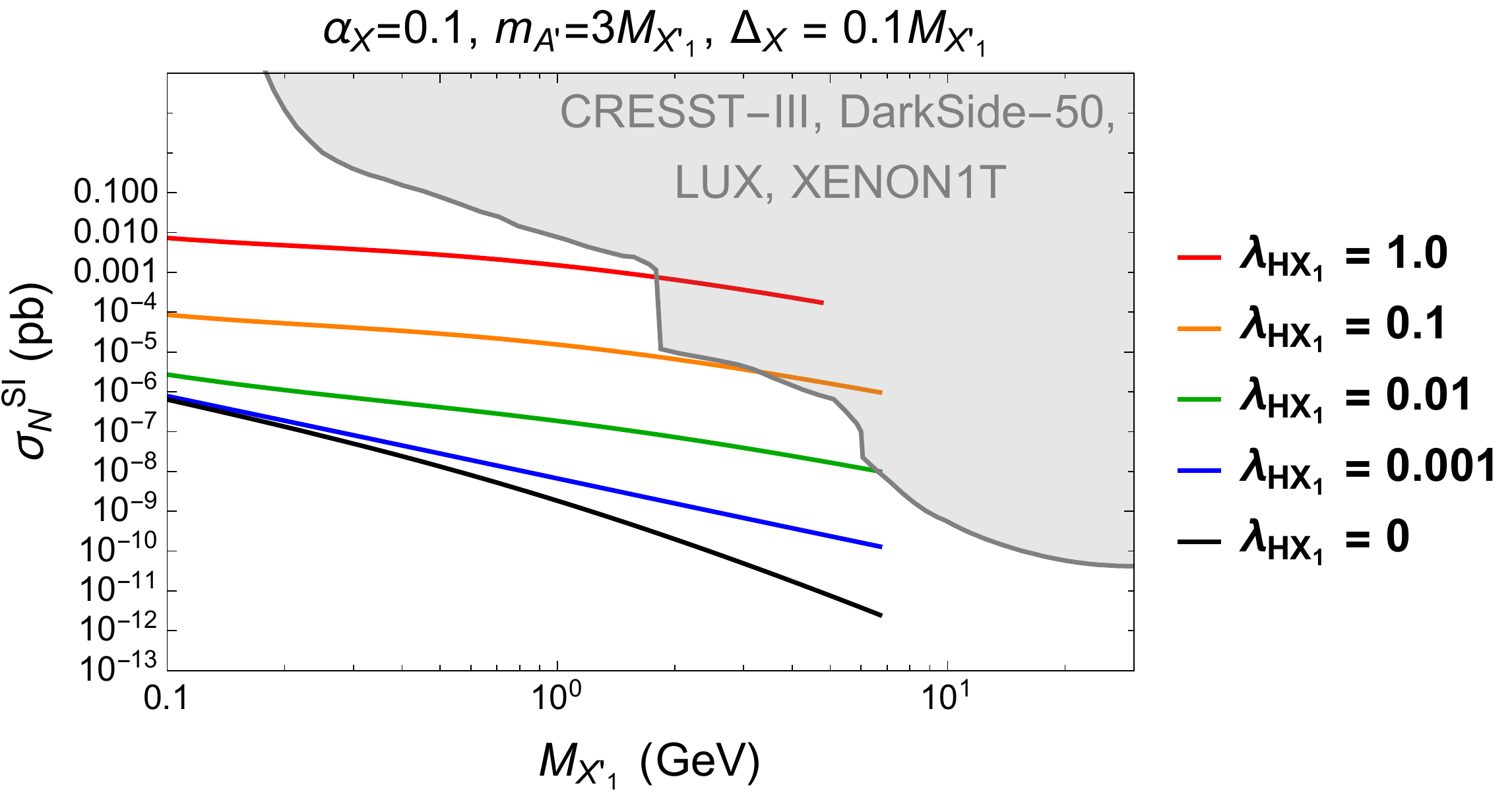} 
\includegraphics[width=3.0in]{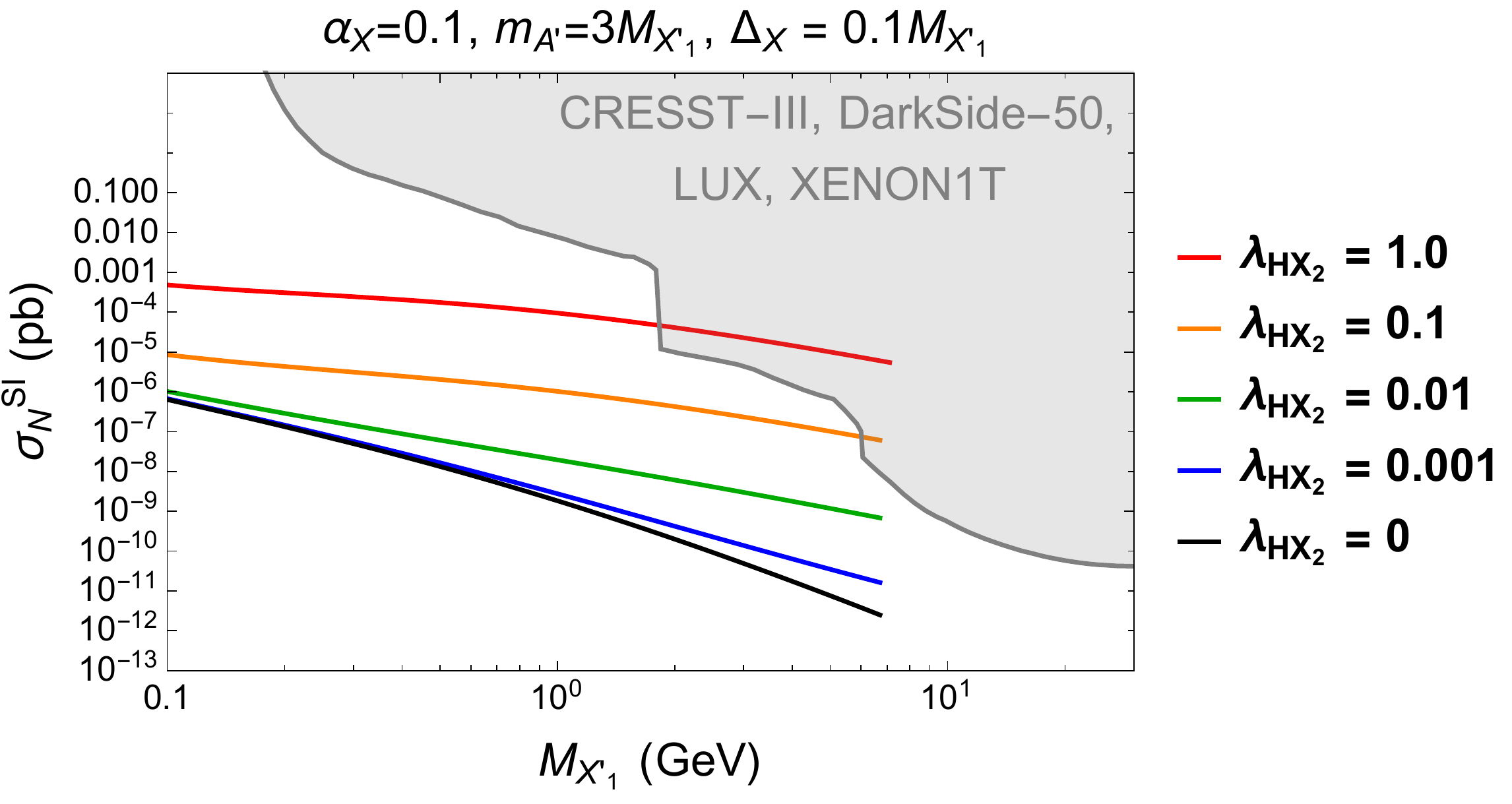}
\includegraphics[width=3.0in]{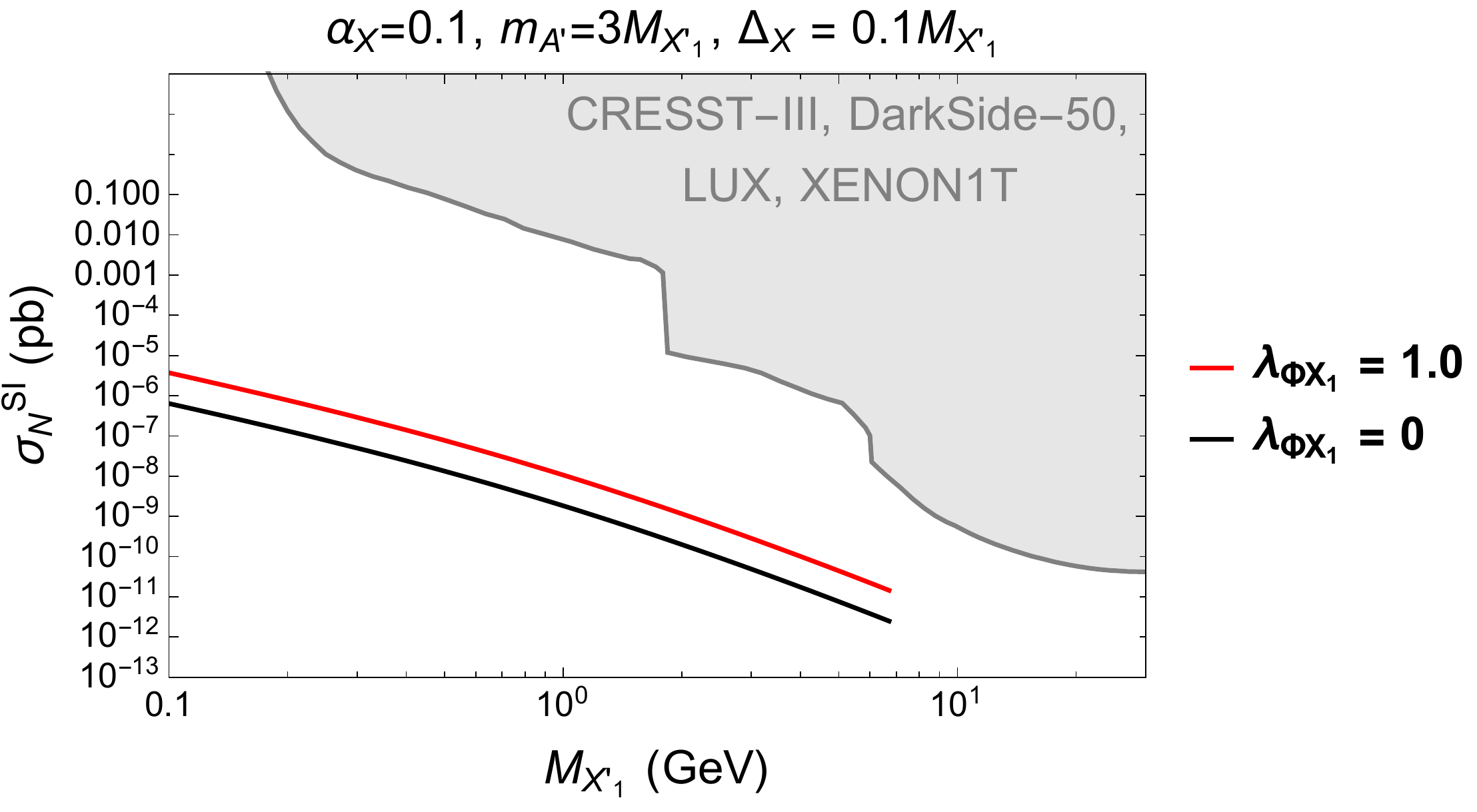}
\includegraphics[width=3.0in]{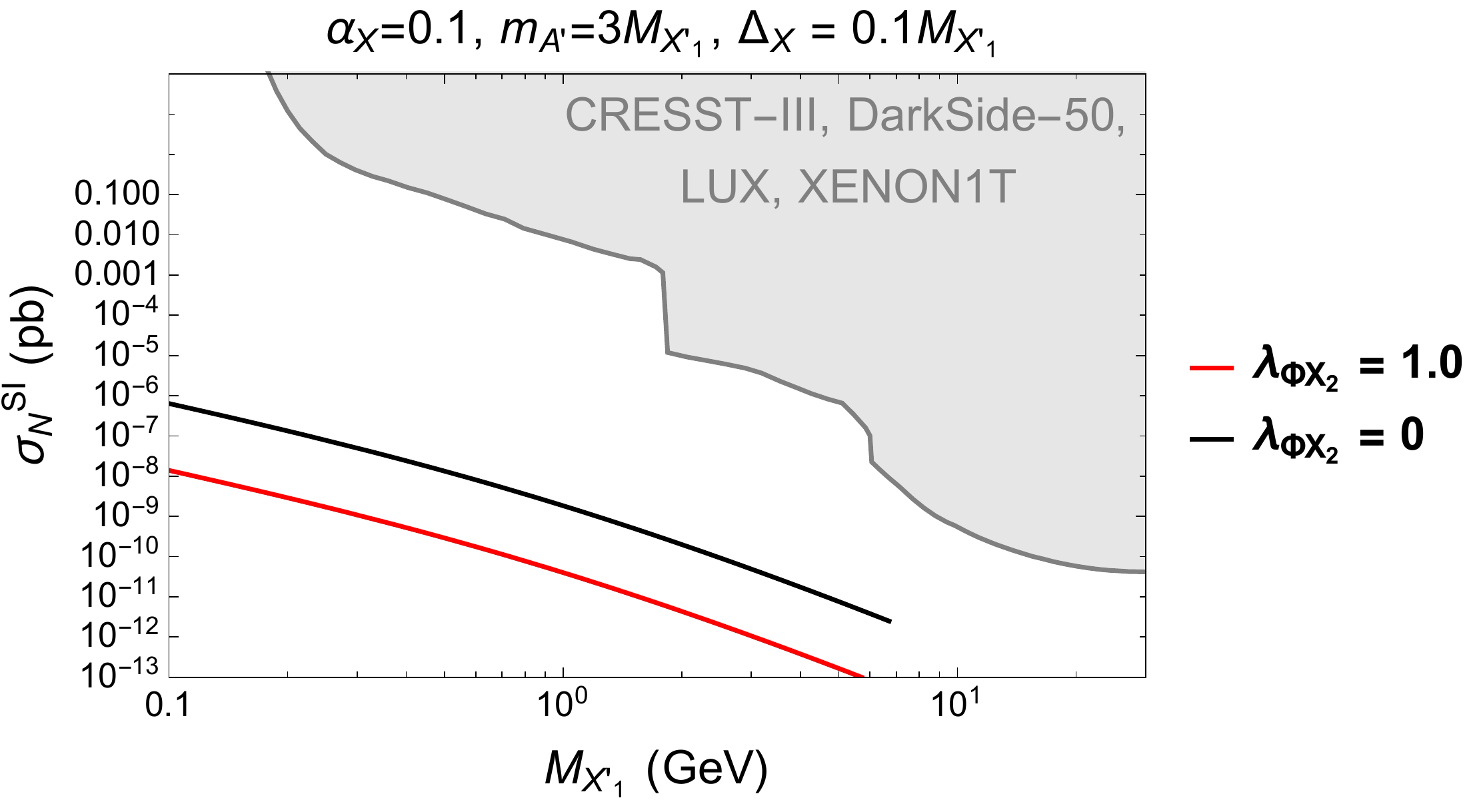}
\caption{ 
{The allowed parameter space with nonzero $\lambda$'s and $\sin\theta_X = 1/\sqrt{1-q_X}$, $\alpha_X = 0.1$, $m_{A'} = 3 M_{X'_1}$, $\Delta_X = 0.1M_{X'_1}$ in the scalar C2CDM models. In the upper-left panel, $(M_{X'_1},\sigma^{SI}_N)$ with varying $\lambda_{HX_1}$. In the upper-right panel, $(M_{X'_1},\sigma^{SI}_N)$ with varying $\lambda_{HX_2}$. In the lower-left panel, $(M_{X'_1},\sigma^{SI}_N)$ with varying $\lambda_{\Phi X_1}$. In the lower-right panel, $(M_{X'_1},\sigma^{SI}_N)$ with varying $\lambda_{\Phi X_2}$. Black solid lines ($\lambda$'s $=0$) are shown for the comparison. The gray regions are excluded by direct detections as explained in the main text. All lines satisfy the observed DM relic abundance.}
}\label{fig:lambda}
\end{figure}

In the main text of the scalar C2CDM models, we made the following assumptions for simplicity:
\begin{equation}
\lambda_{X_1} = \lambda_{X_2} = \lambda_{X_1 X_2} = \lambda_{HX_1} = \lambda_{\Phi X_1} = \lambda_{HX_2} = \lambda_{\Phi X_2} = 0~,
\end{equation} 
in order to make direct comparison with fermionic C2CDM models.
In general, however, these nonzero quartic couplings can affect the DM relic densities and the DM-nucleon scattering cross-sections. In Fig.~\ref{fig:lambda}, we show how nonzero quartic couplings, $\lambda_{HX_1}$, $\lambda_{HX_2}$, $\lambda_{\Phi X_1}$ and $\lambda_{\Phi X_2}$, can modify the DM-nucleon scattering cross-sections. Note all lines in Fig.~\ref{fig:lambda} satisfy 
the observed DM relic abundance.

The results can be summarized as follows.
First, $M_{X^\prime_1}$ cannot exceed about $7 \, {\rm GeV}$ for all lines in Fig.~\ref{fig:lambda} because the upper bound $\epsilon <  0.1$ is considered by Fig.~\ref{fig:Ap_cons}. 
Second, for nonzero $\lambda_{HX_1}$ and $\lambda_{HX_2}$, the DM elastic scattering with nucleon mediated by SM Higgs is enhanced due to the dim-4 operators, $H^\dagger H X^*_1 X_1$ and $H^\dagger H X^*_2 X_2$. 
Third, for nonzero $\lambda_{\Phi X_1}$ and $\lambda_{\Phi X_2}$, the DM elastic scattering with nucleon mediated by dark Higgs is also affected, but the effects from these two parameters are suppressed by the small Higgs mixing angle $\sin \theta_h$. Finally, the effects from nonzero $\lambda_{X_1}$, $\lambda_{X_2}$ and $\lambda_{X_1 X_2}$ to the DM-nucleon scattering cross-sections can be ignored compared to the case of all $\lambda$'s $=0$ because they are important only for the DM self-interactions.

\end{document}